\documentclass[aps, superscriptaddress, 12pt, notitlepage]{revtex4-1}
\usepackage[utf8]{inputenc}
\usepackage[sort&compress]{natbib}
\usepackage{amsmath, graphicx, amsfonts, amssymb, amsthm, subfig, hyperref, booktabs, enumerate, array, footnote, tablefootnote, color, float, tikz, gensymb, algorithm, algorithmicx, wasysym, mathtools, bbm, soul, enumitem}
\newcommand{\x}{\textbf{x}}
\newcommand{\vel}{\textbf{u}}
\newcommand{\svel}{\textbf{v}^s}
\newcommand{\n}{\textbf{n}}
\newcommand{\tang}{\textbf{t}}
\newcommand{\grav}{\textbf{g}}
\newcommand{\zero}{\boldsymbol{0}}
\newcommand{\e}{\textbf{e}}
\newcommand{\G}{\mathcal{S}}
\newcommand{\stress}{\boldsymbol{\Pi}}
\newcommand{\identity}{\textbf{I}}
\newcommand{\projection}{\textbf{P}^s}

\begin{document}

\title{Deformed Liquid Marble Formation: Experiments and Computational Modelling}
\author{Jesse~R.~J.~Pritchard}
\author{Mykyta~V.~Chubynsky}
\affiliation{Mathematics Institute, University of Warwick, Coventry CV4 7AL, United Kingdom}
\author{Jeremy~O.~Marston}
\affiliation{Department of Chemical Engineering, Texas Tech University, Lubbock, Texas 79409, USA}
\author{James~E.~Sprittles}
\affiliation{Mathematics Institute, University of Warwick, Coventry CV4 7AL, United Kingdom}

\date{\today}

\begin{abstract}
The formation of deformed liquid marbles via impact of drops onto powder beds is analysed using experimental and computational modelling approaches.  Experimentally, particular attention is paid to determining a relationship between the maximum contact area of the spreading drops, which determines how much powder the drop's surface is able to harvest, and the drop's surface area when the powder (potentially) encapsulates and then immobilises (`freezes') the surface of the drop to form a liquid marble.  Comparisons between impacts on powder beds to those on rigid and impermeable superhydrophobic substrates show good agreement for a range of parameters and motivate the development of the first mathematical model for the process of liquid marble formation via drop impact.  The model utilises experimentally-determined functions to capture encapsulation and freezing thresholds and accounts for the powder's influence on the drop via a surface viscous mechanism.  Simulations in the volume-of-fluid framework qualitatively recover many features of the experiments and highlight physical effects that should be incorporated into future analyses.
\end{abstract}

\maketitle

\section{Introduction}{\label{SEC:Introduction}}

A `liquid marble' is formed when a liquid drop is completely encapsulated by a shell-like structure, typically consisting of solid particles. This prevents the interior liquid from wetting solids, or coalescing with other volumes of liquid \cite{Quere2001}, so that they are able to efficiently transport small volumes of interior liquid \cite{Quere2001, Mahadevan2001nonstick}.  For example, some of the first liquid marbles analysed, consisting of liquid drops (volume $\sim1-10$mm$^3$) coated with hydrophobic lycopodium grains (characteristic size $\sim 20\mu$m) \cite{Quere2001}, allowed liquids of varying viscosities to be transported along an otherwise wettable solid substrate at high speeds, with little imposed force.

Liquid marbles have undergone significant study, and many viable applications to real-world problems have been proposed. For example, as microscale reactors for instigating chemical reactions \cite{Carter2010pausing, Miao2014catalytic, Wei2016liquid}, microscale bio-reactors for micro-organism cultivation \cite{Tian2013respirable} and human blood typing \cite{Arbatan2012liquid}, contamination detection in liquid and gases \cite{Bormashenko2009revealing, Tian2010gassensing, Fujii2011liquid}, and multi-scale lens production \cite{Shin2019compoundlenses}. An interested reader is directed to review articles describing applications of liquid marbles \cite{Bormashenko2011liquid, Mchale2011liquid, McHale2015, Oliveira2017potential, Nguyen2020liquid} for further information.

\emph{Deformed} liquid marbles are created when particles forming the surface of the marble fully encapsulate it and then approach close-packing, which `freezes' the interface and prevents any further reduction in drop surface area, and thus recapture of a symmetric/spherical drop shape \cite{Marston2012, Marston2013freezing, Supakar2016spreading}. This is in contrast to liquid marbles with sufficiently sparse particle coatings, where the adhered particles can freely rearrange during drop dynamics and (importantly) the surface area reduces to that of a sphere (or the appropriate minimal energy configuration). Notably, the formation and properties of deformed liquid marbles have gained considerably less attention than their un-deformed counterparts and will be the focus of this article.

An efficient method for creating deformed liquid marbles is to impact the carrier liquid as a drop onto a hydrophobic powder bed at sufficiently high speed \cite{Marston2012, Marston2013freezing}. For experiments with an impact speed above a critical threshold, a sufficient quantity of particles become stuck to the drop's surface during its spreading phase so that upon rebound a jammed state is achieved and the drop surface is immobilised (described as `freezing drop oscillations' or an `arrested interface'). Past this threshold, as impact speeds increase the shapes of deformed liquid marbles tend to become more elongated and cylindrical \cite{Marston2012, Supakar2016spreading} until volume loss due to splashing complicates matters.

{\color{black} The conditions for deformed liquid marble formation involve the diameter of the impacting spherical drop $D_0$ \cite{Marston2012, Supakar2016spreading} and the impact speed $U$.  Dimensionlessly, in the regime we study, where viscous and gravitational effects have a negligible influence on the spreading dynamics (see \cite{Supakar2016spreading} for estimates), this means the transition from spherical to deformed drops is characterised solely by a sufficiently large Weber number $\text{We} = \rho~U^2~D_0/\sigma$ (liquid density $\rho$, surface tension of the (clean) liquid-gas interface $\sigma$), with experiments showing a critical value of We$^*$ $\approx$ 60-70 \cite{Supakar2016spreading}.}  Physically, larger $U$ results in a greater spreading diameter for the drop and hence the opportunity for the drop to harvest enough powder particles to freeze the interface upon rebound, before the drop returns to a sphere.  It is also possible that whilst this picture gives qualitative insight into the process, further details, such as how much powder is actually required to encapsulate and then freeze an interface, remain unknown \cite{Supakar2016spreading, Mozhi2019predictive} and are difficult to address experimentally due to the small spatio-temporal scales and complex dynamics. 

At the heart of the problem is an understanding of how the maximum contact area of the drop, a circular footprint of diameter $D_{\text{contact}}$, relates to the surface area of the drop (i) at encapsulation and (ii) at the point of interfacial freezing.  Estimates for the encapsulation have been considered, that $D_{\text{contact}} \ge 2 D_0$ \cite{Mozhi2019predictive}, based on the contact area $\pi D_{\text{contact}}^2/4$ being larger than the sphere's surface area $\pi D_0^2$.  However, the ejection of satellite drops upon rebound \cite{Marston2012}, which reduce the drop's mass/surface-area, makes simple estimates inaccurate and has thus far only been accounted for empirically \cite{Supakar2016spreading} to yield a criterion $D_{\text{contact}} > 1.67 D_0$. Notably, there has been no detailed study of how the drop becomes encapsulated and, most importantly, how this relates to the freezing of the interface, which does not in general occur at the same time.

Despite exciting experimental discoveries, there remains little understanding on the interplay of powder and droplet dynamics which creates deformed liquid marbles.  This will be addressed here using new targeted experiments on liquid marble formation and exploiting their results to guide the first computational model for this phenomenon.  This model will capture the influence of the powder on the drop through a continuum method that treats the powder as a surface phase.  Specifically, following the Boussinesq-Scriven approach \cite{Scriven}, we will show that a dilatational surface viscosity can be used to describe many of the features of deformed liquid marble formation. This should allow us to predict, for example, what impact speeds are required to create deformed marbles.

In \S\ref{SEC:Experimental_Methods}-\ref{SEC:Experimental_Results} experiments on liquid marble formation focus on the relationship between $D_{\text{contact}}$ with the surface area of the drop at both encapsulation and deformed liquid marble formation. Additionally, we compare impacts on powder beds to those on a flat rigid impermeable superhydrophobic substrate, as their similarity has important modelling consequences.  Then, in \S\ref{SEC:Modelling}, we construct a computational model for liquid marble formation via drop impact, with the influence of dissipative surface effects \cite{Reynaert2007interfacial, Zang2010viscoelastic, Kotula2012probing, Bykov2014dilational, Barman2016role} incorporated. In \S\ref{SEC:Simulations_Model_Refinement} we discuss the numerical implementation of the problem into open-source volume-of-fluid software and how modelling parameters are refined following preliminary numerical simulations. In \S\ref{SEC:Simulation_Results}, simulation results are compared to experiments before, in \S\ref{SEC:Discussion}, a discussion of results is given, with avenues for future work highlighted.

\section{Experimental Methods}{\label{SEC:Experimental_Methods}}
The particles used to form the powder beds in experiments are soda lime solid glass microspheres (Cospheric, USA) that arrive pre-treated with a hydrophobic nano-coating.  {\color{black}The effect of creating a porous substrate composed of hydrophobic microspheres gives the substrate superhydrophobic properties, and so we refer to it as a `superhydrodphobic powder bed'}. The stated particle density from the supplier is 2.45g/cc, and after conducting a time-of-flight measurement using an API Aerosizer\textsuperscript{\textregistered} particle size analyzer, the geometric mean particle diameter is found to be approximately 24$\mu$m, with tenth- and ninetieth-percentiles of 12$\mu$m and 46$\mu$m, respectively. For the rigid impermeable superhydrophobic substrates, glass microscope slides are washed with acetone and de-ionised water and then immersed for a few seconds in a liquid suspension (Glaco Mirror Coat, Soft 99 Co., Japan) containing silica nanoparticles, before being left to air dry for five minutes. The glass slides are then heat cured on a hot plate at 120$\degree$C for five minutes, after which the slides exhibit superhydrophobic properties due to the aggregation of nanoparticles across their surfaces. The liquid used in all experiments is de-ionised water, the relevant properties of which (at $20^\circ$C) are the density $\rho = 9.98 \times 10^2$ kg/m$^3$, dynamic viscosity $\mu = 1.00\times 10^{-3}$ kg/m$\cdot$s, and surface tension $\sigma = 72.75 \times 10^{-3}$ kg/s$^2$.

\begin{figure}
	\begin{center}
        	\includegraphics[width=\textwidth*19/20]{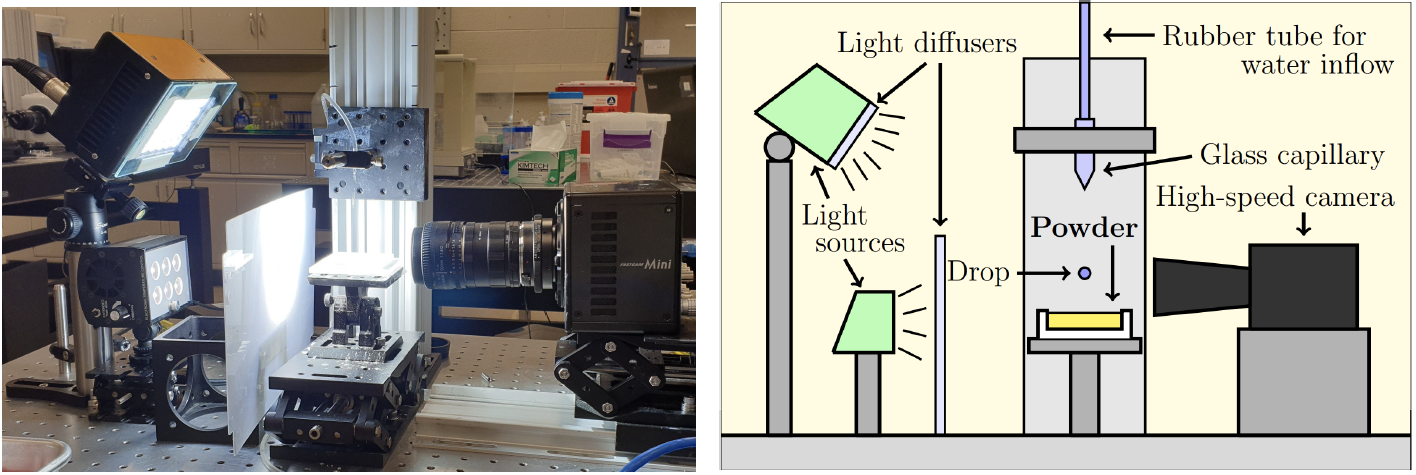} 	
	\end{center}
    \caption{Photograph and diagram of the experimental setup for drop impacts onto a superhydrophobic powder bed.}
    {\label{FIG:Experiment_Setup}}
\end{figure}

The experimental setup for impacts onto superhydrophobic powder beds is shown in Figure \ref{FIG:Experiment_Setup}. Water drops are created by feeding de-ionised water through a rubber tube connecting a 30ml syringe to a glass capillary that is clamped into position at a given height. The thin end of the capillary (where the drops emerge) is hydrophobised to ensure that drops do not wet the outside of the capillary when forced out, meaning that drop size is determined by the size of the capillary opening (along with the value of liquid surface tension). For impacts onto rigid superhydrophobic substrates, the powder bed is replaced by a glass slide which has been hydrophobised using the above process.

{\color{black} The impact speed, $U$, is varied by changing the height $H$ of the capillary opening with respect to the substrate, so that upon neglecting the effects of air resistance the impact speeds are based on free-fall calculations}. The range of drop heights and impact speeds for our experiments are $4.6 \le H \le 466.7$mm and $0.30 \le U \le 3.02$m/s, respectively.  Drop diameters $D_0$ are 1.99mm for impacts on powder substrates and 1.94mm on hydrophobised glass. {\color{black} The drop diameters were calculated from images of the drop in free fall (taking an average of the vertical and horizontal values), with errors on the order of the pixel size (see below) and the measurements seen to be repeatable across experiments.} The experiments are illuminated using two light sources, each placed behind light diffusers to maximise contrast of the impacting drops with the background for the sake of visual analysis. Experiments are recorded at 4000 frames per second using a monochrome FASTCAM Mini UX100 high-speed camera, and Photron FASTCAM Viewer software is used for visual data extraction. The error in measurements of visual data is expected to be on the order of 1 pixel whose width is 22$\mu$m for powder experiments and 24$\mu$m for glass. Notably, each impact (on powder or hydrophobised glass) is on a new/fresh part of the substrate. A description of the image analysis techniques used to identify the maximum spreading diameter and an approximation of drop surface area from experimental images {\color{black} is provided} in Appendix \ref{APP:Image_Analysis}.

\section{Experimental Results}{\label{SEC:Experimental_Results}}

Within this section, we provide a detailed analysis of powder bed experiments before considering impacts on the rigid and impermeable hydrophobised glass. A direct measurement of the maximum contact diameter is impossible from the experiments, as the position of the contact line cannot be accurately identified due to the high contact angles present. Instead, the maximum spread of the impacting drop on the substrates is characterised by the (dimensionless) spreading factor $\gamma = D_{max} / D_0$, where $D_{max}$ is the maximum drop diameter visible during experiments (what you would see from a bird's eye view).

\begin{figure}
	\begin{center}
	\begin{center}
        	\includegraphics[width=\textwidth*19/20]{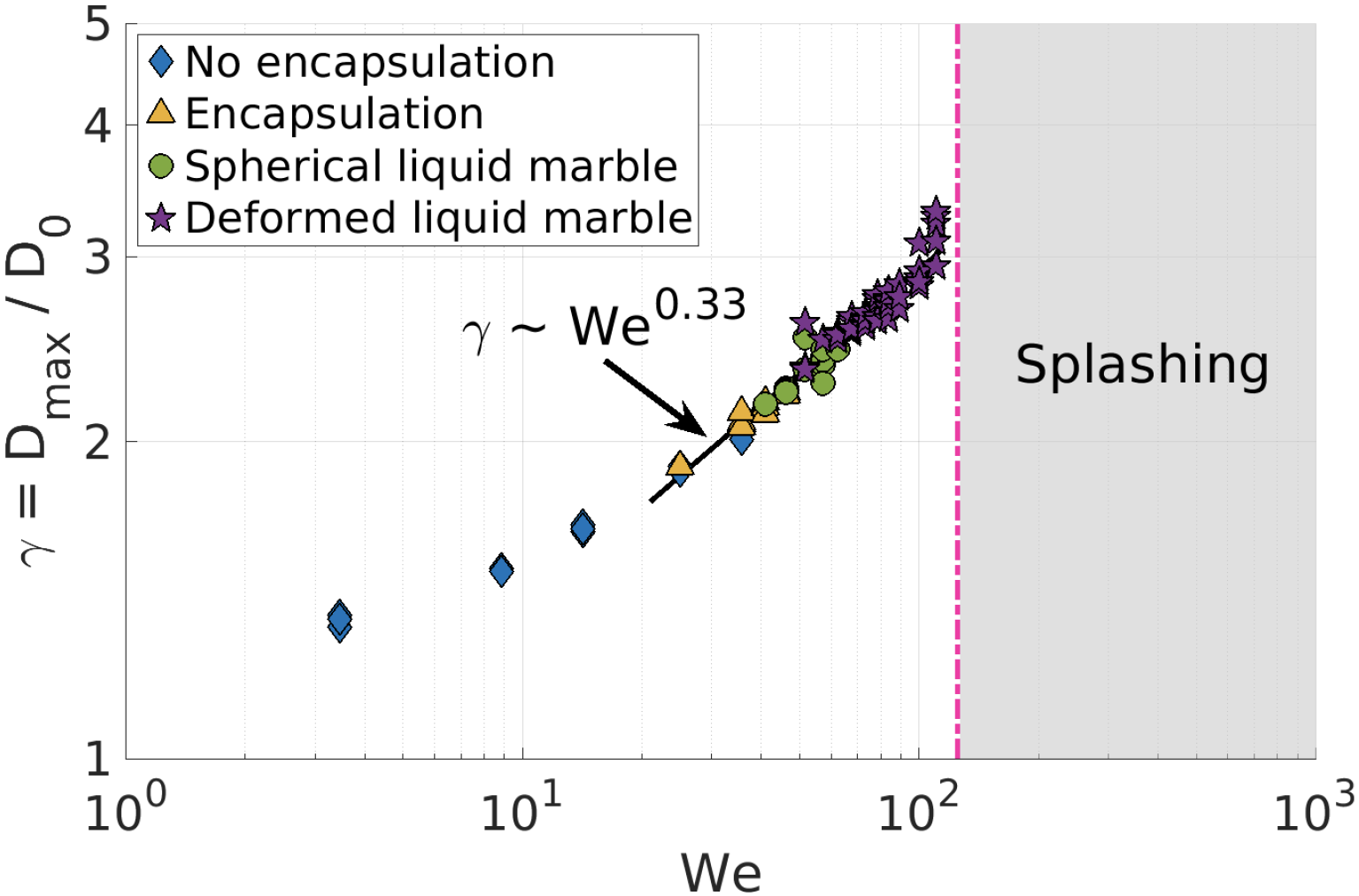} 	
	\end{center}
	\end{center}
    \caption{Spreading factor versus impact Weber number for drop impact onto a superhydrophobic powder bed. A power law approximation is provided for $\text{We} > 20$.}
    {\label{FIG:DropSpread_PowderImpacts}}
\end{figure}

In Figure \ref{FIG:DropSpread_PowderImpacts}, $\gamma$ is plotted against the impact Weber number and follows an approximate power-law of We$^{0.33}$ for We $> 20$, in line with the literature of drop impact onto superhydrophobic powder beds \cite{Supakar2016spreading, Mozhi2019predictive}. The different drop regimes are highlighted, which are similar to those given in existing research \cite{Marston2012, Supakar2016spreading}, except that we make the crucial distinction between drops becoming encapsulated and being `liquid marbles'. In particular, the term `liquid marble' is reserved for cases in which the drop encapsulation leads to (i) a fast damping of oscillations and a quick return to a spherical shape (`spherical liquid marbles') or (ii) the formation of a deformed (i.e. non-spherical) liquid marble.

\begin{figure}
	\begin{center}
        	\includegraphics[width=\textwidth*19/20]{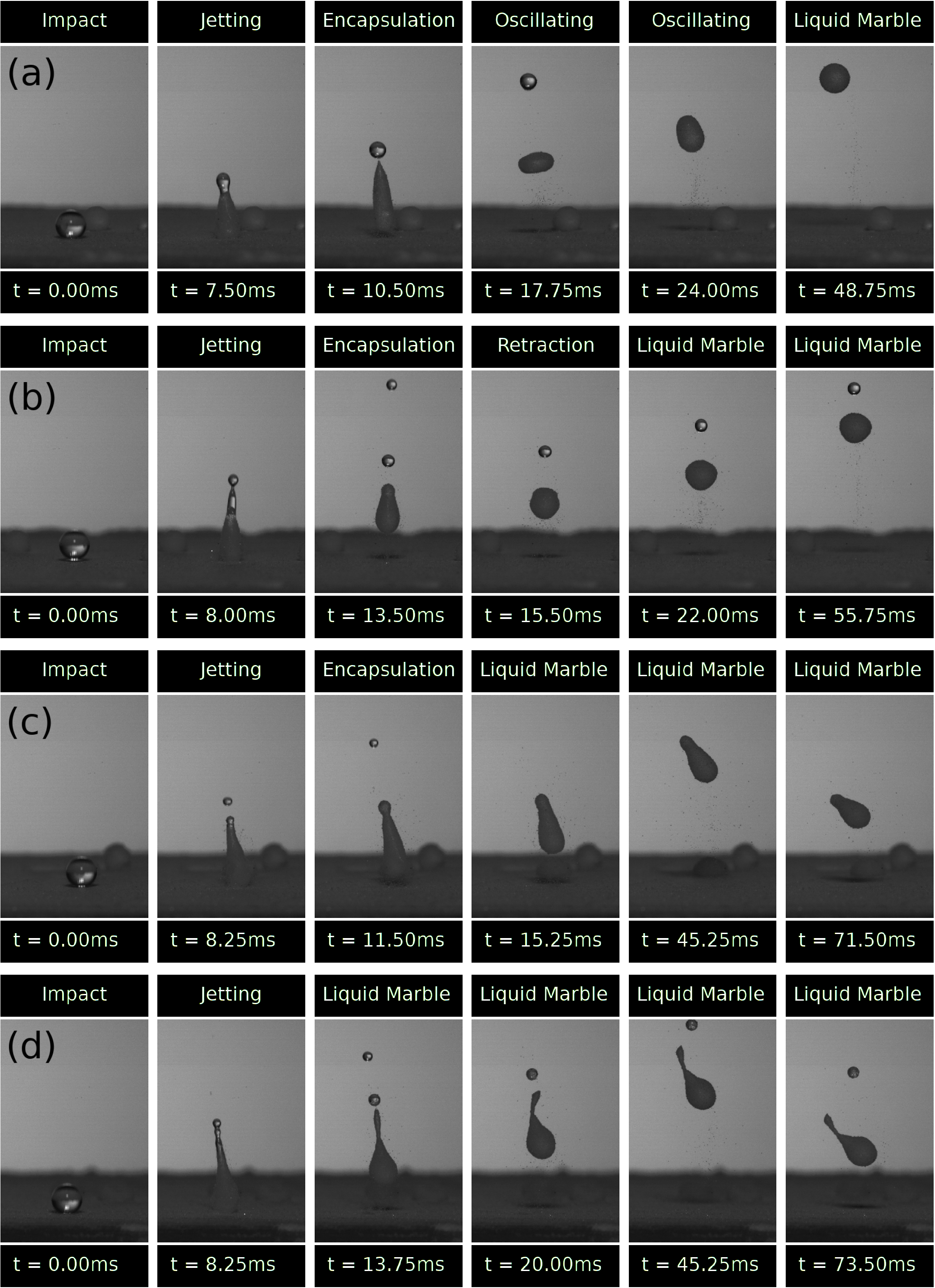} 	
	\end{center}
	\caption{Liquid marbles formed with different impact Weber numbers. (a) We = 46 spherical marble, (b) We = 52 slightly deformed liquid marble, (c) We = 73 elongated deformed marble, (d) We = 78 elongated deformed marble maintaining the same shape as at the moment of encapsulation.}
    {\label{FIG:Experiments_PowderBed}}
\end{figure}

In Figure \ref{FIG:DropSpread_PowderImpacts}, we introduce the following regimes:
\begin{itemize}[noitemsep,nolistsep]
 \item \emph{No Encapsulation} -- is observed for lower impact Weber numbers, with drops only attaining a partial coating of powder on their interfaces. 
\item \emph{Encapsulation} --  is seen for a narrow range of impact Weber numbers where the drop can become encapsulated, but no liquid marble (neither spherical nor deformed) is created, as the drops continue to oscillate vigorously throughout their rebound motion until they fall back under gravity to the substrate.  
\item \emph{Spherical Marbles} -- are seen to form for We $> 40$, see Figure \ref{FIG:Experiments_PowderBed}(a); in this case drop oscillations quickly decay, and the drop attains a spherical shape prior to landing back on the powder bed, often seen occuring before reaching the apex of its rebound flight. 
\item \emph{Deformed Marbles} -- appear from We $> 52$, see Figure \ref{FIG:Experiments_PowderBed}(b), with slightly deformed shapes maintained following apparent jamming of the particle-laden interface.  As the impact Weber number is increased, the arrested shape of the deformed liquid marble becomes more elongated as seen for We = 73 in Figure \ref{FIG:Experiments_PowderBed}(c).  Note that the shape at encapsulation is close to that of the liquid marble itself and upon further increases the shape at encapsulation is the same as the marble, as shown for We = 78 in Figure \ref{FIG:Experiments_PowderBed}(d). These values are in line with previous experimental analyses \cite{Marston2012, Supakar2016spreading}.
\item \emph{Splashing} -- for We $>$ 115, splashing occurs soon after the moment of impact on the powder bed; deformed liquid marbles are still produced, but these cases are beyond the scope of this article and so are neglected.
\end{itemize}

\subsection{Dependence of Marble Formation on Maximum Spread/Contact}{\label{SEC:MarbleFormation_as_DropSpread}}

For the first time, we investigate how the change in the maximum contact diameter affects the two critical moments in liquid marble formation; namely drop encapsulation (full coverage of the drop interface in adhered particles) and the jamming of the drop interface. In the remainder of this article, we refer to the jamming of particles at the interface as `interfacial freezing' or `drop freezing', following the initial description of deformed liquid marble formation as ``freezing drop oscillations" \cite{Marston2012}.

\subsubsection{Converting Spreading Areas to Contact Areas}

\begin{figure}
	\begin{center}
        	\includegraphics[width=\textwidth*19/20]{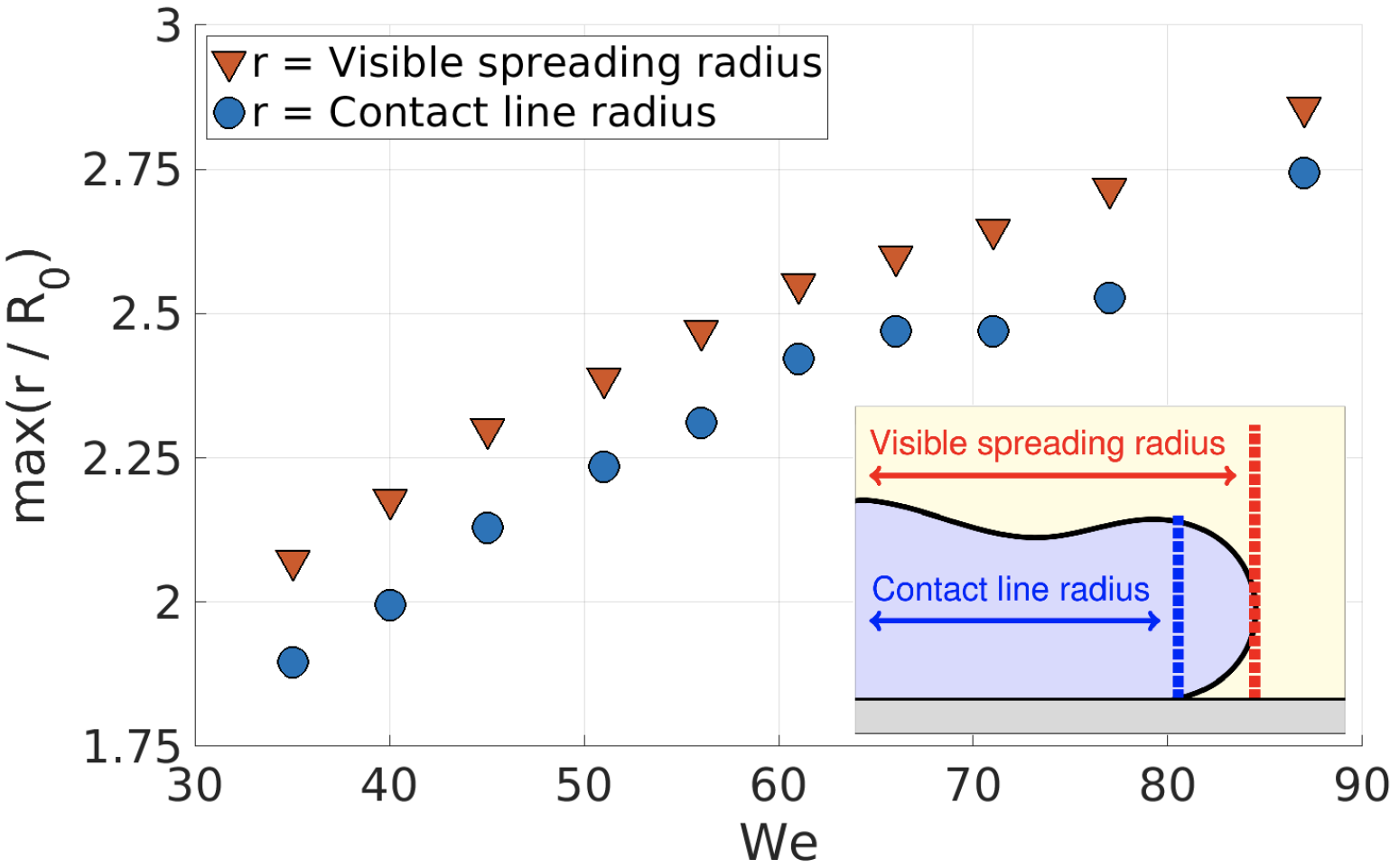} 	
	\end{center}
    \caption{Maximum visible drop radius and maximum contact line radius, divided by the initial spherical drop radius $R_0$, both taken at the moment of maximum spread for drop impact simulations, as a function of the impact Weber number.  {\color{black}The inset shows how these two quantities are defined}}
    \label{FIG:LM_Radius_Alpha_a}
\end{figure}

As is the case for the maximum contact diameter, the obscuring of the contact line position also prevents a direct measurement of the maximum contact area in experiments, so instead the `spreading area' is measured; defined as the area of a disc with radius $D_{max}$ so that $A_{spread} = \pi D_{max}^2 / 4$. Using simulations (cf. \S\ref{SEC:Simulations_Model_Refinement}) of drop impact (see Figure \ref{FIG:LM_Radius_Alpha_a}), we find that the maximum radial extension of the contact line and the maximum visible drop radius differ by approximately a constant (given dimensionally as 0.16 times the initial spherical radius); a relationship we use to convert our experimental data concerning the spreading area to the more physically relevant contact area $A_{contact}$. Henceforth, we will only consider the physically-relevant contact areas and remember that implicitly we have determined these from spreading areas.

\subsubsection{Encapsulation}
Before encapsulation, the drop interface is split into a clean region and a powder coated region, with encapsulation occurring when the surface area of the clean region goes to zero and the powder coats the entire interface. At maximum spread, the powder coated region of the drop interface is exactly the part of the drop in contact with the powder bed, and it is after the drop starts to retract that we see the powder spread along the liquid-gas interface, {\color{black} although experimentally we cannot see how this evolution relates to the liquid's velocity distribution}. Here, we consider what relationship (if any) exists between the surface area of all powder coated regions (this includes partially coated satellite drops) at the moment of encapsulation, and the maximum contact area between the drop and the powder bed.

\begin{figure}
	\begin{center}
        	\includegraphics[width=\textwidth*19/20]{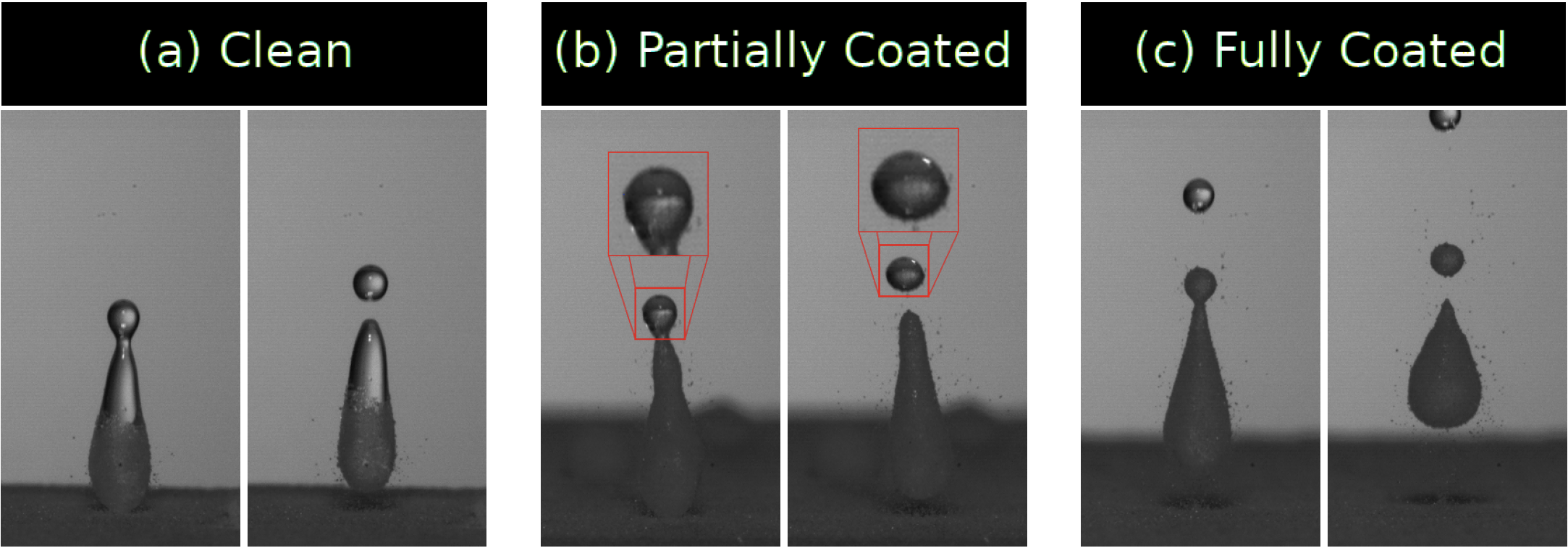} 	
	\end{center}    \caption{Drop impact experiments on a superhydrophobic powder bed showing the ejection of a satellite drop with a liquid-gas interface that is: (a) clean, (b) partially coated, (c) fully coated. Cases (b) and (c) reduce the total mass of powder adsorbed to the liquid-gas interface of the primary drop.}
    \label{FIG:Pinchoff}
\end{figure}

Notably, a direct comparison between the surface area of the primary drop at encapsulation and $A_{contact}$ cannot be made, because although there are some cases where ejected satellite drops are clean (Figure \ref{FIG:Pinchoff}(a)), in others they remove powder to initiate encapsulation (Figure \ref{FIG:Pinchoff}(b)). In the latter case, the mass of powder on the primary drop interface is no longer equal to what it was at maximum spread.

To capture these scenarios, we consider the total surface area the powder covers across all drops when the primary drop becomes encapsulated $A_{encap}^{(0)}$, which we call the `initial encapsulation area'.  It is the sum of the surface areas of the primary drop and the powder coated surface area of any partially coated satellite which, if formed, necessarily triggered the encapsulation of the primary drop. Then, by defining 
\begin{equation}{\label{EQN:Experiments_alpha_encap}}
    \alpha_{encap} = \frac{A_{encap}^{(0)}}{A_{contact}},
\end{equation}
we have a (dimensionless) quantity that describes the extent of the powder coverage across all regions at encapsulation as a proportion of the maximum contact area. Naively, we may expect that $\alpha_{encap}$ takes a constant value, meaning that the drop surface area covered by powder at the moment of encapsulation is some fixed proportion of the contact area (as implicitly assumed in previous works); however, we will later see that this is not quite the case.

\subsubsection{Interfacial Freezing}

Our aim here is to construct a similar expression to (\ref{EQN:Experiments_alpha_encap}) for freezing via an `initial freezing area', $A_{freeze}^{(0)}$.  Freezing of a drop interface is the result of a tight packing of adsorbed powder preventing further surface area reduction, so that it is best described in terms of powder `surface concentration' $c$, a mass per unit area. However, as we can only measure the surface area of powder regions experimentally, we make the simplest assumption that powder concentration is spatially constant, but can vary in time $c = c(t)$. Then, powder concentration is inversely proportional to the (measurable) surface area ($A$) of the powder coated region $c \propto A^{-1}$, so that $c$ rising to a critical packing threshold is equivalent to $A$ reducing to a critical area threshold.  The coefficient of proportionality is the total mass of adsorbed powder $M$ so that $c = M/A$.

If there is no reduction in mass of adsorbed powder on the primary drop via satellite ejections, then the interface should freeze once the surface area of the drop falls to $A_{freeze}^{(0)} = M_{contact}/c_{freeze}$, where $M_{contact}$ is the total powder mass adsorbed through interaction with the powder bed, and $c_{freeze}$ represents the critical packing threshold necessary for interfacial freezing (particle jamming). We would like to compare this initial freezing area across different drop impact experiments, but given that many experiments \emph{do} exhibit pinch-off events that reduce the adsorbed powder mass (both as a catalyst for encapsulation as in Figure \ref{FIG:Pinchoff}(b), or while the drop is already encapsulated as in Figure \ref{FIG:Pinchoff}(c)), this requires some careful attention.

The total surface area of the primary drop is clearly not continuous through a pinch-off event, nor is the total mass of adsorbed powder on the primary drop \emph{if} the satellite drop has some powder coating on its interface. On the other hand, powder concentration on coated regions \emph{is} continuous through all pinch-off events.  For a drop that experiences multiple satellite drop ejections, we can write $A^{(0)}_{freeze}$ as a function of the surface areas of the multiple droplets, which are all measurable in experiments, see Appendix \ref{APP:Initial_Freezing_Area} for the expression. Therefore, we can now define the number
\begin{equation}{\label{EQN:Experiments_alpha_freeze}}
    \alpha_{freeze} = \frac{A_{freeze}^{(0)}}{A_{contact}},
\end{equation}
which describes the critical area threshold for freezing as a proportion of the maximum contact area.

\subsubsection{Determination of Initial Encapsulation Areas and Initial Freezing Areas: Finding the $\alpha$'s}

\begin{figure}
	\begin{center}
        	\includegraphics[width=\textwidth*19/20]{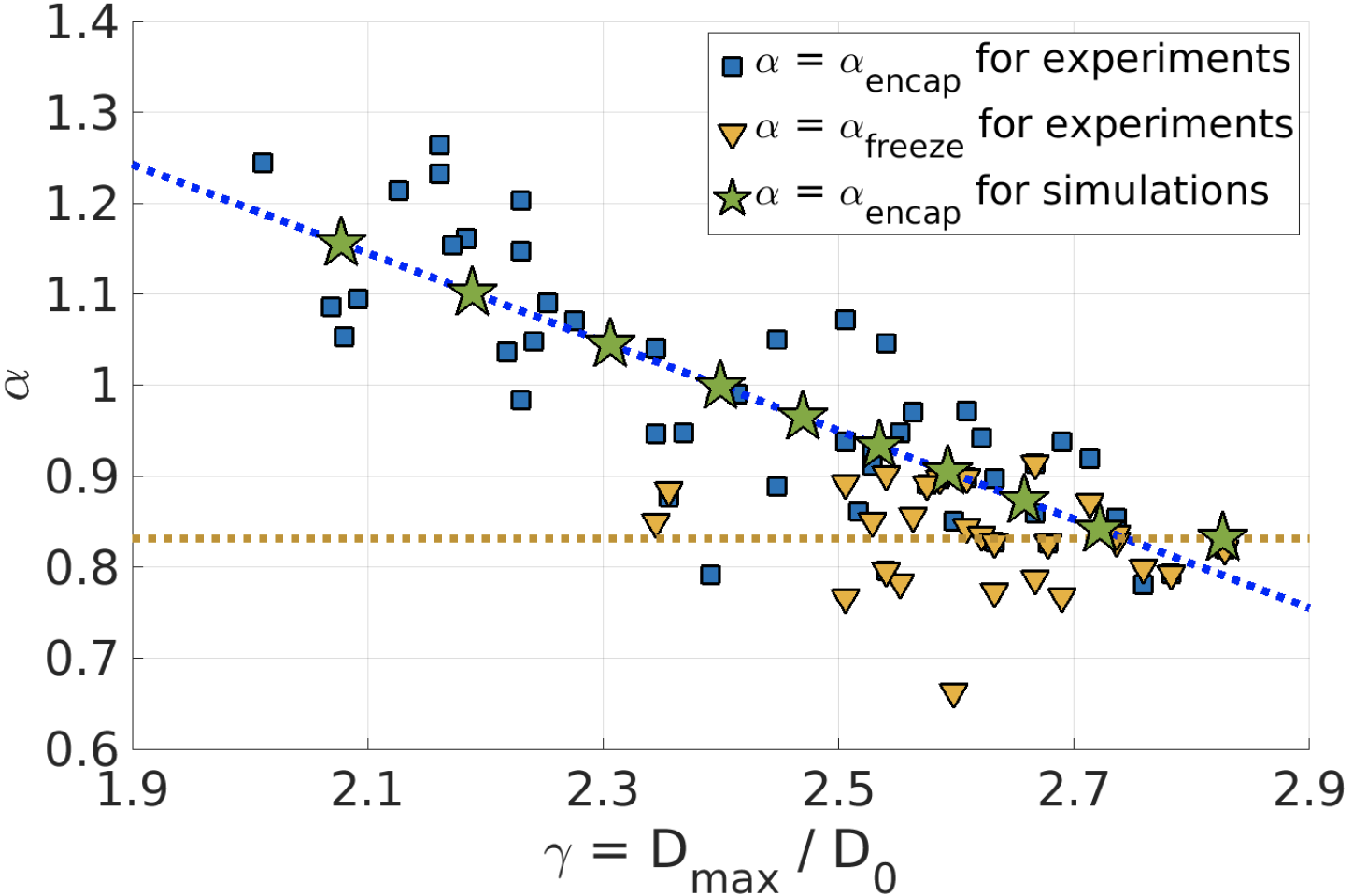} 	
	\end{center}    \caption{Dependence of $\alpha_{encap}$ and $\alpha_{freeze}$ on the spreading ratio $\gamma$. A linear fit is given for the $\alpha_{encap}$ data points, and a constant fit for the deformed $\alpha_{freeze}$ data points. Values of $\alpha_{encap}$ are given for simulations of drop impact with We = 35, 40, 45, 51, 56, 61, 66, 71, 77, 87, shown in terms of $\gamma$, which increases monotonically with increasing We. For We = 87, $\alpha_{encap}$ is set equal to $\alpha_{freeze}$.}
    \label{FIG:LM_Radius_Alpha}
\end{figure}

The dependence of $\alpha_{encap}$ and $\alpha_{freeze}$ on the spreading factor $\gamma$ is shown in Figure \ref{FIG:LM_Radius_Alpha} using data gathered from the powder bed experiments. The vertical axis is denoted by $\alpha$ that represents the surface area of a drop at the given critical event, divided by the maximum contact area. 

For encapsulation, $\alpha_{encap}$ clearly decreases (on average) as the spreading factor increases, with a linear approximation given by $\alpha_{encap}(\gamma) = -0.49 \gamma + 2.17$; in other words, drops which spread further encapsulate at lower surface areas as a proportion of their maximum contact area.  The physical consequences of this result will be discussed further in \S\ref{later}, but its investigation is beyond the main scope of this work, where we will directly use this phenomenological expression in developing a model.

For deformed liquid marble formation, data points are clustered around an average value of $\alpha = 0.83$, showing that if a rebounding drop can reduce to a surface area of 0.83 times its maximum contact area then it is likely to become a deformed liquid marble, and this is almost guaranteed for $\gamma \ge 2.5$. The downward trend of $\alpha_{encap}$ for increasing $\gamma$ leads to a decreasing interval of surface areas between encapsulation and interfacial freezing, culminating in several points where $\alpha_{encap} = \alpha_{freeze}$, that is, where the two events occur simultaneously, as described earlier on.  

The $\alpha$ functions are used later in our computational model and marked on Figure \ref{FIG:LM_Radius_Alpha} are the values at which simulations are performed. Note that the value of $\alpha_{encap}$ for the simulation data point with the largest $\gamma$ lies on the fitted line for $\alpha_{freeze}$ as otherwise the linear fit for $\alpha_{encap}$ suggests that freezing precedes encapsulation, a point we will comment on later.

\newpage

\subsection{Comparison to Rigid Substrate Impacts}

\begin{figure}
	\begin{center}
        	\includegraphics[width=\textwidth*19/20]{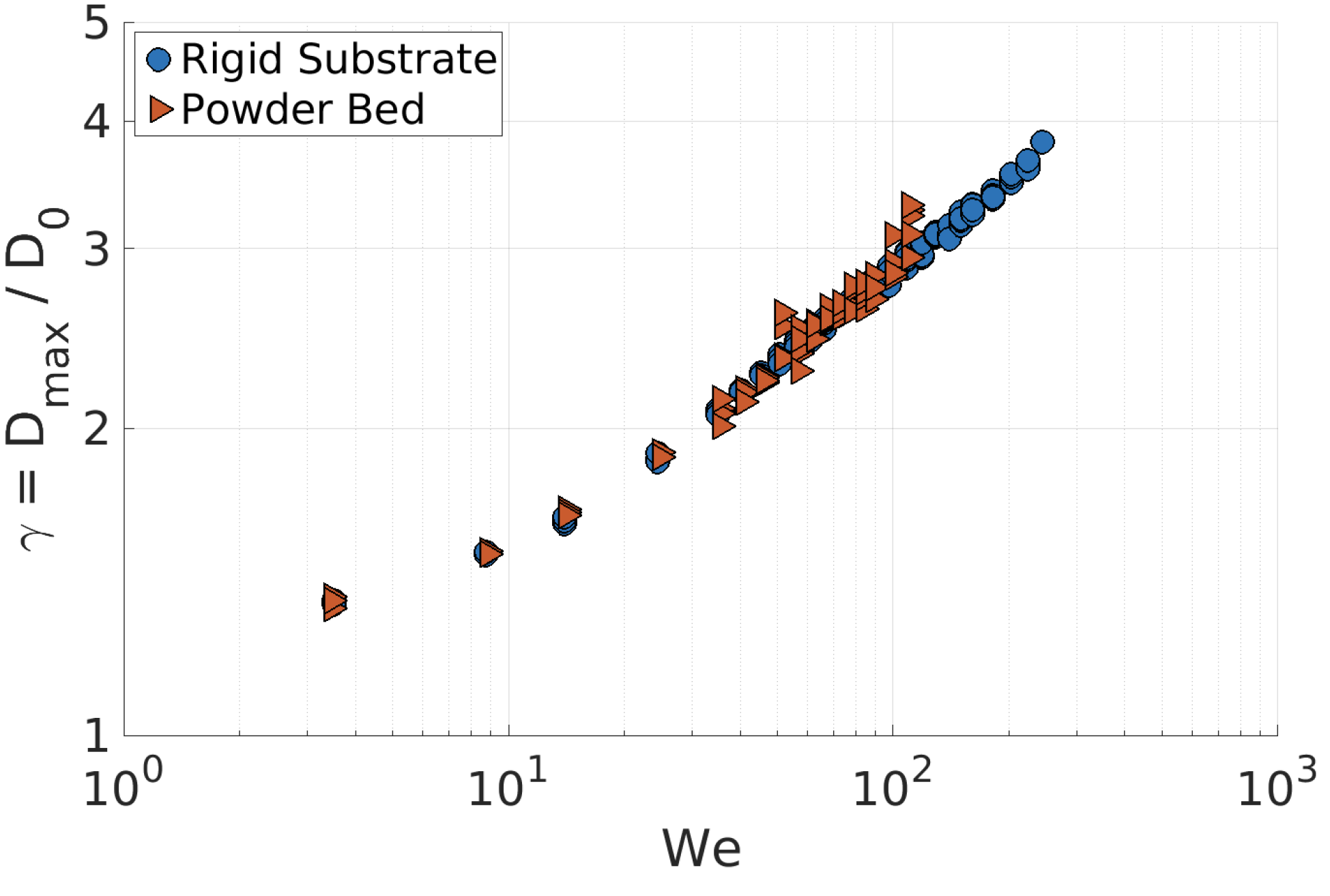} 	
	\end{center}
    \caption{Spreading factor versus impact Weber number for drop impacts onto a superhydrophobic powder bed and rigid impermeable superhydrophobic substrate.}
    {\label{FIG:DropSpread_AllImpacts}}
\end{figure}

It is interesting to consider the relationship between impacts on powder beds with the more commonly considered case of rigid superhydrophobic substrates, and their similarity will have important consequences for the development of a computational model. Provided in Figure \ref{FIG:DropSpread_AllImpacts} is a combined plot of the spreading factor against impact Weber numbers for both sets of experiments; we see that as well as having similar scaling behaviour, there is good agreement on the actual value of the spreading factor across a wide range of We (albeit with a wider variance for the powder bed impacts). The similarity between the experiments on different substrates prompts further investigation. Although a particular value for the spreading factor is not entirely indicative of rebound dynamics, the data presented in Figure \ref{FIG:DropSpread_AllImpacts} suggest that for a range of impact Weber numbers, impacts on superhydrophobic powder beds may be reasonably approximated by impacts on rigid impermeable superhydrophobic substrates.

\begin{figure}
	\begin{center}
        	\includegraphics[width=\textwidth*19/20]{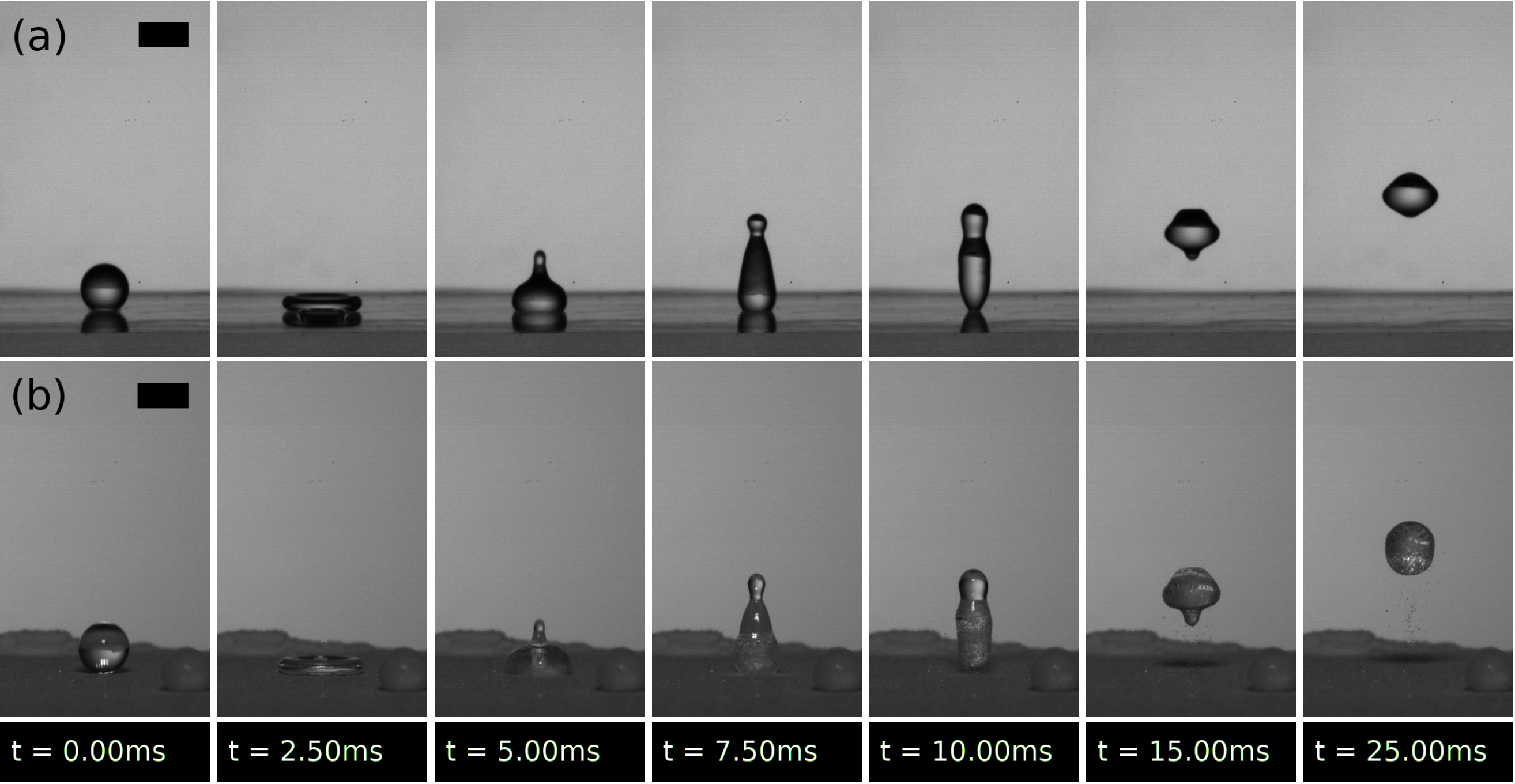} 	
	\end{center}
    \caption{Comparison of drop impacts on (a) a rigid impermeable superhydrophobic substrate, and (b) a superhydrophobic powder bed. (a) $D_0 = 1.94$mm and $\text{We} = 14$, with spreading factor $\gamma = 1.62$, (b) $D_0 = 1.99$mm and $\text{We} = 15$, with spreading factor $\gamma = 1.67$. Scale bar has a width of 2mm.}
    {\label{FIG:Comparison_h3}}
\end{figure}

\begin{figure}
	\begin{center}
        	\includegraphics[width=\textwidth*19/20]{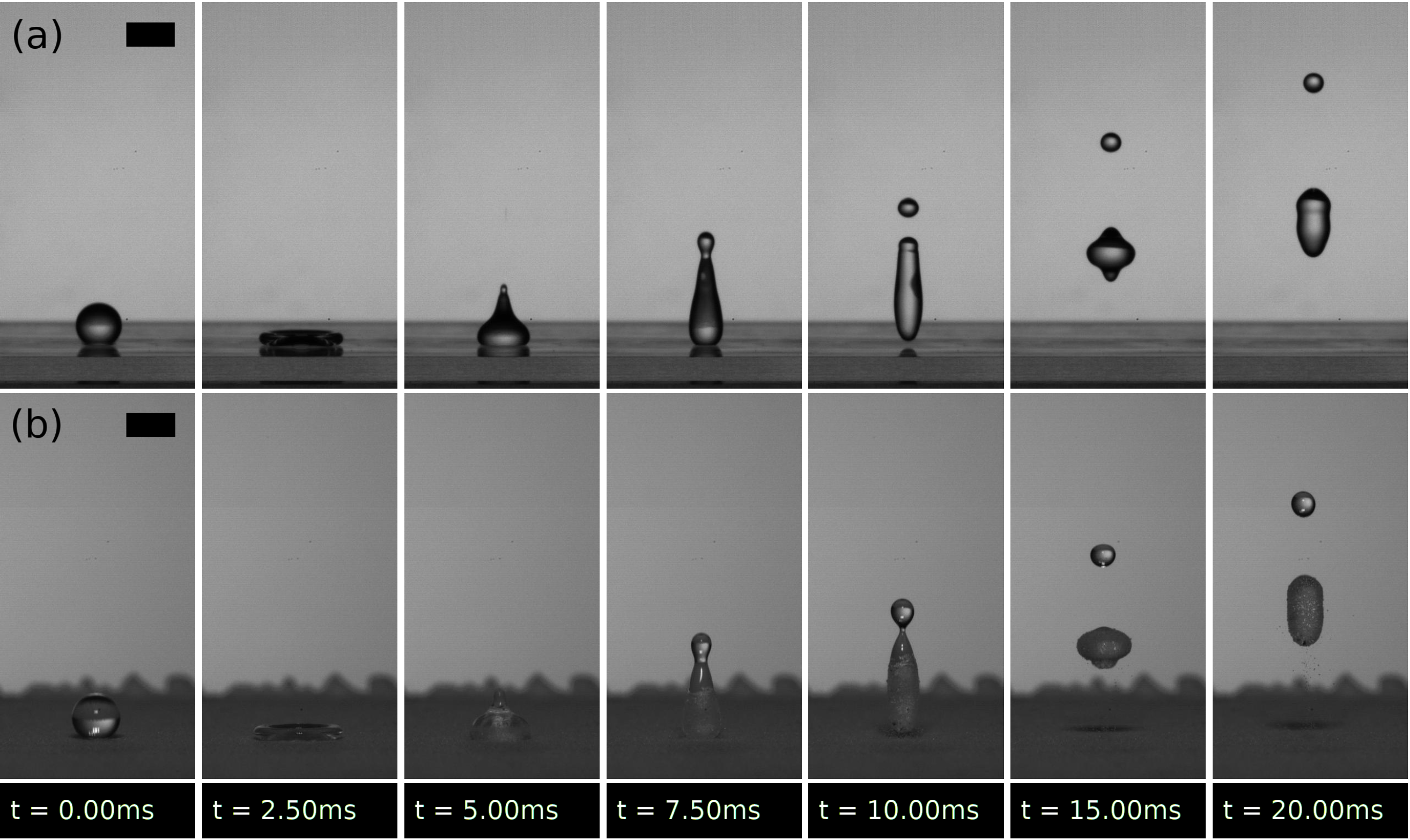} 	
	\end{center}
    \caption{Comparison of drop impacts on (a) a rigid impermeable superhydrophobic substrate, and (b) a superhydrophobic powder bed. (a) $D_0 = 1.94$mm and $\text{We} = 24$, with spreading factor $\gamma = 1.87$, (b) $D_0 = 1.99$mm and $\text{We} = 25$, with spreading factor $\gamma = 1.87$. Scale bar has a width of 2mm.}
    {\label{FIG:Comparison_h4}}
\end{figure}

Provided in Figures \ref{FIG:Comparison_h3}-\ref{FIG:Comparison_h4} are visual comparisons between characteristic impacts on rigid impermeable superhydrophobic substrates and superhydrophobic powder beds for similar impact Weber numbers ranging from We $= 14$ to $25$. Through spreading, retraction, and for some time after rebound, the drop shapes exhibit striking similarities, for example the intricate `spinning-tops' at t = 15.00ms in Figures \ref{FIG:Comparison_h3}-\ref{FIG:Comparison_h4} appears for both types of substrate. 

\begin{figure}[H]
	\begin{center}
        	\includegraphics[width=\textwidth*19/20]{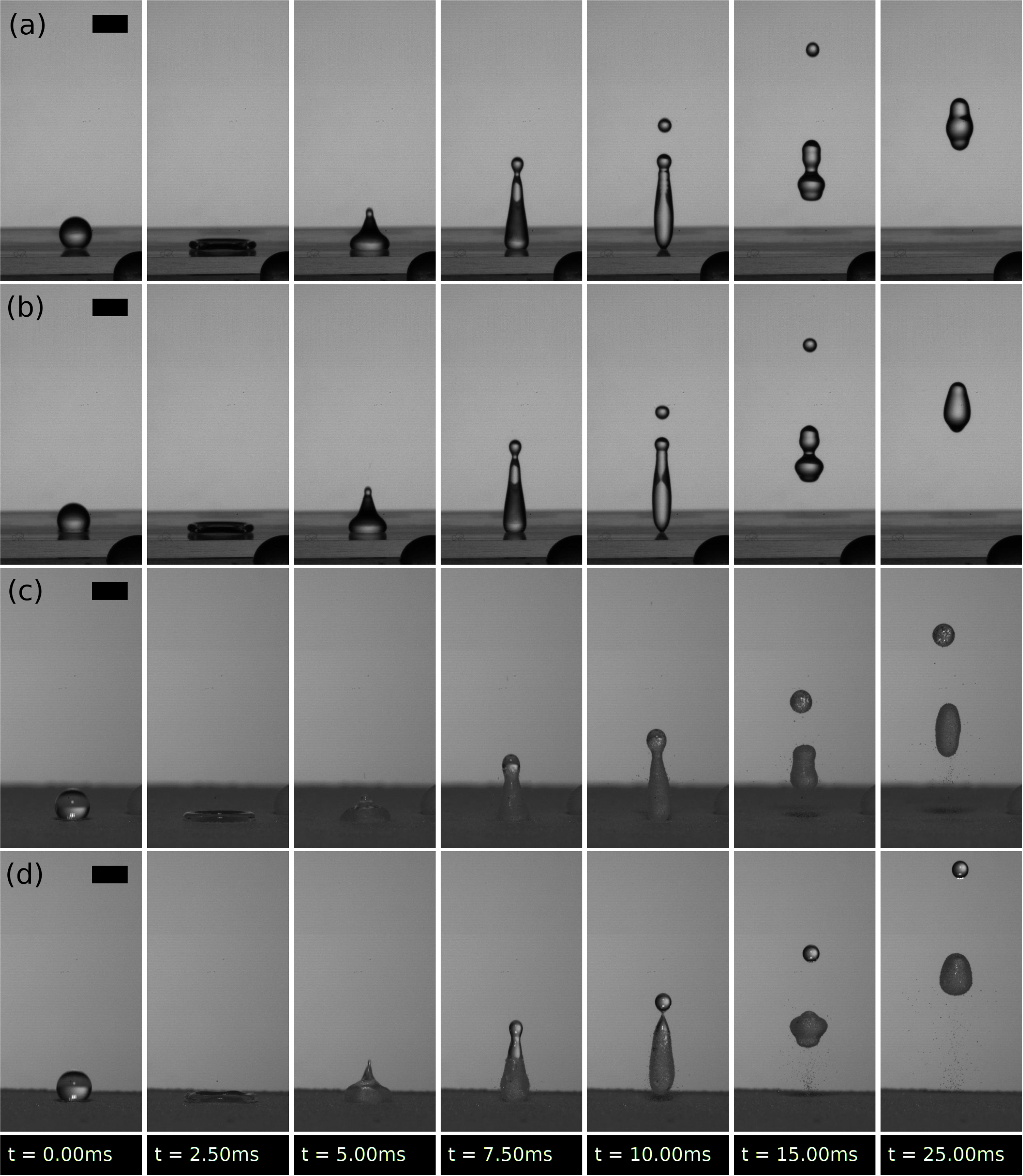} 	
	\end{center}
    \caption{Comparison of drop impacts on (a-b) a rigid impermeable superhydrophobic substrate, and (c-d) a superhydrophobic powder bed. (a-b) $D_0 = 1.94$mm and $\text{We} = 35$, with spreading factor $\gamma = 2.06$, (c) $D_0 = 1.99$mm and $\text{We} = 36$, with spreading factor $\gamma = 2.07$, (d) $D_0 = 1.99$mm and $\text{We} = 36$, with spreading factor $\gamma = 2.14$. Scale bar has a width of 2mm.}
    {\label{FIG:Comparison_h5}}
\end{figure}

For higher impact Weber numbers (We $\ge 36$), there is still a similarity between the two types of experiments, as portrayed in Figure \ref{FIG:Comparison_h5}.  Notably, before this regime, impacts onto powder beds were reproducible, however we see here that this is not the case; at time $t = 7.50$ms the difference in powder coverage between the (c) and (d) is quite considerable, indicating that interactions between the drop and the powder bed have started to become very important. By contrast, impacts on rigid impermeable superhydrophobic substrates continue to be reproducible, as seen in (a) and (b), and their dynamics are close to the (now range of) powder-bed impacts.  

We have seen here that for a range of Weber numbers, drop shapes exhibited in experiments for impacts onto superhydrophobic powder beds are largely similar to those seen after impacts on rigid impermeable superhydrophobic substrates. In particular, what is key for our model, is that the overall dynamics, i.e. the drop shapes, are similar, despite some small alterations in time scales which become more pronounced at higher $\text{We}$. Therefore, for our model, as a first approximation, we will consider the powder bed to be rigid and impermeable, to circumvent the need to model the powder bed deformation.

\section{A Model for Liquid Marble Formation}{\label{SEC:Modelling}}

The liquid marble formation process is complex, due to the high deformation and non-standard surface dynamics, so our focus here will be on developing the \emph{simplest} model for the process, with numerous potential extensions and improvements discussed in \S\ref{SEC:Discussion}.

\subsection{Problem Formulation}

Consider an initially spherical drop with radius $R_0=D_0/2$ of an incompressible Newtonian fluid with (constant) density $\rho$, and dynamic viscosity $\mu$. Denote the domain contained within the drop by $\Omega$, and the bulk velocity by $\vel$. The smooth interface $\G_{LG}$, separating the bulk liquid drop phase from an exterior gas phase, is characterised by the (constant) surface tension $\sigma$, along with surface shear and surface dilatational viscosity coefficients $\mu^s$ and $\lambda^s$, respectively, which can become non-zero due to the presence of adsorbed particles on the drop interface. We denote the velocity of the liquid-gas interface $\G_{LG}$ (also known as the surface velocity) by $\svel$; importantly this may differ from the value of bulk velocity taken at the interface due to the presence of adsorbed particles. 

Beneath the drop there exists an isotropic superhydrophobic powder bed, modelled in our case as a flat rigid impermeable superhydrophobic substrate. We denote by $\G_{LS}$ (when it exists) the interface between the drop and the solid substrate.

We adopt a cylindrical co-ordinate system $(r, \theta, z)$, where $r$ and $z$ denote the radial  and axial coordinates, respectively, and $\theta$ denotes the azimuthal angle about the $z$-axis, and we assume axisymmetry, in line with experimental observations. The $z$-axis (where $r = 0$) is chosen to coincide with the vertical axis (perpendicular to the substrate) that passes through the centre of mass of the (spherical) drop prior to impact, so that the initial drop velocity is $-U\e_z$, and the $r$-axis (where $z = 0$) is chosen to coincide with the surface of the flat rigid substrate.

The reduction via axisymmetry means that the drop's liquid-gas interface $\G_{LG}$ can be considered, in the $(r, z)$-plane, as a one dimensional boundary curve with a single parameter that spans the interface. This parameter is chosen to be the arclength, $s$, and ranges from $s = 0$ where the curve `first' intersects the axis of symmetry (without loss of generality choose the intersection with greatest $z$ coordinate), to $s = L > 0$ at the second intersection, or at the contact line where it meets the liquid-solid interface $\G_{LS}$, if the latter interface exists.

\subsubsection{Modelling the Powder Dynamics}{\label{SEC:ProblemDynamics}}

We treat the adhered particle coating on the drop surface as a continuum and model the effects of the (continuum) particle coating using surface viscosity, which is incorporated using the Boussinesq-Scriven approach \cite{Scriven} that effectively considers the surface dynamics as Newtonian. Other approaches are possible, including the use of discrete particle-based models \cite{Frijters2012deformation, Gunther2013lattice, Kruger2013numerical, Harting2015lattice}, but the continuum approach is a computationally efficient and popular starting place for surface dynamics. This introduces surface shear and surface dilatational viscous coefficients, which we consider to be negligible for a clean interface and otherwise depend on the concentration of adhered particles $c$ within the region under consideration through a constitutive relation.  Our focus will be on dilatational effects (so $\mu^s = 0$) as, particularly for higher impact Weber number cases, surface area change acts as a catalyst for interfacial freezing, while there is no clear surface shearing. 

The influence of the powder bed on drop spreading is incorporated into our model via a slip length that affects the drop's flow in a small boundary layer close to the substrate, as this is a standard way of modelling drop spreading on (macroscopically) flat superhydrophobic surfaces \cite{Rothstein2010slip}.

\subsection{Surface Equations}

We begin with development of the surface equations and conclude with the bulk equations, an unorthodox approach which is motivated by considerable simplification to the surface equations when an inviscid flow assumption is made.  Notably, the Reynolds numbers $\text{Re} = \rho~U D_0/\mu$ based on impact speed are typically large, with encapsulation occurring for Re $> 1600$, so that inviscid flow appears reasonable.  However, whilst this is the case for the rebound dynamics, during the impact phase strong boundary layer effects at the liquid-solid interface and the formation of thin films (particularly near the apex) create regions where the local Reynolds number is much smaller and viscous fluid dynamics is necessary, as we shall later describe.

\subsubsection{Liquid-Gas Surface Equations}{\label{SEC:LiquidGas_SurfaceEquations}}

Let $f(\x, t) = 0$ be the implicit equation of the liquid-gas interface $\G_{LG}$ such that $f < 0$ for $\x \in \Omega$ (liquid) and $f > 0$ for $\x \in \Omega_G$ (gas), where $\x$ and $t$ denote space and time, respectively. Consider the unit normal $\n = \nabla f / \left|\nabla f\right|$ pointing from the liquid phase, so that assuming no mass flux through the interface the kinematic equation gives
\begin{equation}{\label{EQN:UpdatedKinematicCondition}}
	\frac{\partial f}{\partial t} + (\vel \cdot \n)|\nabla f| = 0 \quad \text{on} \ \G_{LG}.
\end{equation}

Neglecting surface body forces and surface inertia \cite{InterfacialTransport}, the balance of forces at the surface gives:
\begin{equation}{\label{EQN:SurfaceStressBalance}}
	\n \cdot \left( \stress - \stress_G \right) = \nabla_s \cdot \stress^s \quad \text{on} \ \G_{LG}.
\end{equation}
where $\stress$ and $\stress_G$ are the bulk stress tensors of the liquid and gas phases, respectively, and $\nabla_s = \projection \cdot \nabla$ is the surface divergence operator, for surface projection tensor $\projection = \identity - \n\n^T$. The surface stress tensor $\stress^s$ is given by the Boussinesq-Scriven constitutive equation \cite{Scriven} with zero surface shear viscosity (i.e. $\mu^s = 0$):
\begin{equation}{\label{EQN:ReducedBoussinesq-Scriven}}
    \stress^s = \left( \sigma + \lambda^s \nabla_s \cdot \svel \right) \projection.
\end{equation}

\subsubsection{Boundary Conditions at the Liquid-Solid Interface}{\label{SEC:LS_Surf_Equations}}
We incorporate the effects of substrate permeability for the powder bed, modelled in our case as rigid and \emph{impermeable}, via an impermeable Beavers-Joseph boundary condition \cite{beavers_joseph_1967} at an effective flat liquid-solid interface:
\begin{equation}{\label{EQN:BeaversJoseph}}
    u_r = \frac{K^{1/2}}{\alpha_{BJ}}\frac{\partial u_r}{\partial z}\quad\hbox{and} \quad  u_z = 0 \qquad \text{at} \ z = 0,
\end{equation}
where $K$ is the permeability of the substrate, $\alpha_{BJ}$ is the (dimensionless) Beavers-Joseph coefficient, and $u_r \equiv \vel \cdot \e_r$ is the bulk radial velocity. Given that permeability $K$ scales with area, specifically the area of pores on a surface, and that the `pores' of a powder bed are on the order of particle size, we suppose that $K = d_p^2$, where $d_p$ is the mean particle diameter and assume that the Beavers-Joseph coefficient $\alpha_{BJ}$ is unity, for sake of argument.

\subsubsection{Boundary Conditions for the Surface Equations}

At the axis of symmetry, we require on $\G_{LG}$ that $\n \cdot \e_r = 0$ at $r = 0$ which ensures the liquid-gas interface $\G_{LG}$ remains smooth across the axis. When the liquid-solid interface $\G_{LS}$ does \emph{not} exist, this applies at the apex and bottom of the drop. When the drop is in contact with the solid substrate (so $\G_{LS}$ \emph{does} exist), this continues to apply at the drop apex, but at the other end of the liquid-gas interface (now at the contact line), we prescribe a contact angle $\theta_c$.  In the interests of simplicity, we assume a constant contact angle and a value of $\theta_c=160^\circ$, which matches experimental data well and is thus used across all simulations. Notably, conditions on $\svel$ would usually be required, but our forthcoming derivation will show these surface effects manifest themselves through an effective surface tension, so this is not required, as we only require that the tangential component of $\svel$ is bounded.

\subsection{Development of a Simplified Model for the Surface Dynamics}
Given the high Reynolds numbers encountered, and the fact that the most interesting surface dynamics occur upon rebound, where boundary layers at solid surfaces are not present, we will see that it is useful to consider the consequences of assuming inviscid flow, where $\stress = -p\identity$ and where pressure $p$ is taken relative to its constant atmospheric value. Then, combining the standard constitutive equation for the bulk stress tensor and the surface stress tensor in (\ref{EQN:ReducedBoussinesq-Scriven}), with the conservation of momentum equation for the liquid-gas interface (\ref{EQN:SurfaceStressBalance}), expanding out and splitting into normal and tangential components to the interface gives
\begin{equation}{\label{EQN:ExpandedSurfaceStress}}
    -\left( \sigma + \lambda^s \nabla_s \cdot \svel \right) \left( \nabla_s \cdot \n\right)\n = -p\n \quad \hbox{and} \quad \nabla_s \left( \sigma + \lambda^s \nabla_s \cdot \svel \right) = \zero \quad \text{on} \ \G_{LG},
\end{equation}
which we will now simplify. 

For the tangential projection, recall that $\sigma$ is constant (so $\nabla_s \sigma = \zero$ on $\G_{LG}$), and integrate the tangential component of (\ref{EQN:ExpandedSurfaceStress}) over the entire liquid-gas interface to obtain
\begin{equation}{\label{EQN:IntroductionOfW(t)}}
    \lambda^s \nabla_s \cdot \svel = W(t) \quad \text{on} \ \G_{LG},
\end{equation}
where $W(t)$ is spatially constant along $\G_{LG}$ but time dependent. The normal projection (\ref{EQN:ExpandedSurfaceStress}) has the surface viscous term $\lambda^s \nabla_s \cdot \svel$ lying within parentheses with the liquid surface tension, and so will affect the flow in the same manner; hence we can think of this combined term as an \emph{effective surface tension};
\begin{equation}{\label{EQN:EffectiveSurfaceTension_Dimensional}}
    \sigma_{\text{eff}}(t) \equiv \sigma + \lambda^s \nabla_s \cdot \svel = \sigma + W(t),
\end{equation}
which is also dependent on time only. Using this effective surface tension, the normal component of (\ref{EQN:ExpandedSurfaceStress}) is simplified to
\begin{equation}{\label{EQN:Normal_Component_Momentum}}
    p = \sigma_{\text{eff}}(t)\left(\nabla_s \cdot \n\right) \quad \text{on} \ \G_{LG}.
\end{equation}

The unexpected consequence of assuming inviscid bulk flow is that the (dilatational) surface viscous effects manifest themselves everywhere on the interface simultaneously via an effective surface tension, independent of how large the powder concentration (and so $\lambda^s$) is in any particular region (see below). The consequences of this model for an effective surface tension are considered below for different scenarios.

\subsubsection{Partially Coated Liquid-Gas Interface}

For a liquid-gas interface $\G_{LG}$ that has only a partial coating of powder (so the drop is not yet encapsulated), we must have $\lambda^s = 0$ in all regions with \emph{no} powder coating. Hence $W(t) = \lambda^s \nabla_s \cdot \svel = 0$ in these regions also, and as $W(t)$ is spatially independent, we must have that $W(t) = 0$ along the \emph{entire} liquid-gas interface, so \emph{no} surface viscous effects {\color{black} are} felt anywhere.  As encapsulation is always observed \emph{after} the drop has rebounded in experiments, the same is assumed in our simulations, and so the above finding ensures that there are \emph{no} surface viscous effects when the drop is in contact with the solid substrate.

During rebound, given that $W(t) = 0$ everywhere on $\G_{LG}$, we must have $\nabla_s \cdot \svel = 0$ within powder coated regions (where $\lambda^s \neq 0$), so that the surface area of the powder coated region remains unchanged.  In other words, when there is a clean region of the interface to move into, the powder acts as an incompressible (surface-wise) shell.  This means that knowing the surface area of the entire liquid-gas interface and the (fixed) surface area of the powder region is sufficient for tracking the point on the interface separating the clean and powder regions (and in particular, determining when the drop is encapsulated). 

\subsubsection{Fully Coated Liquid-Gas Interface}

For the case of a fully coated (encapsulated) drop, $\lambda^s \neq 0$ everywhere and consequently $W(t)$ needs to be determined. Dividing both sides of equation (\ref{EQN:IntroductionOfW(t)}) by $\lambda^s$, integrating over the liquid-gas interface $\G_{LG}$, rearranging and using $\svel \cdot \n=\vel \cdot \n$, we obtain:
\begin{equation}{\label{WInt}}
	W(t) = \frac{\int_{\G_{LG}} \nabla_s \cdot \svel \ \text{d}S}{\int_{\G_{LG}} 1/\lambda^s \ \text{d}S},\quad\hbox{where}\quad \nabla_s \cdot \svel = \frac{1}{r}\frac{\partial}{\partial s}(r \svel \cdot \underline{\textbf{t}}) + (\nabla_s \cdot \n)(\vel \cdot \n),
\end{equation}
d$S$ denotes an area element of the interface $\G_{LG}$, $\tang$ is the tangent vector to $\G_{LG}$ in the ($r, z$)-plane, and the axisymmetric form of the operators can be found in \cite{pritchard_phd}.

Transforming the surface integrals in (\ref{WInt}) into line integrals via axisymmetry and recognising that $r = 0$ and $\svel \cdot \tang$ is bounded when $s = 0$ and $s = L$ (both at the axis of symmetry), we obtain the following expression for $W(t)$:
\begin{equation}{\label{WEquation}}
	W(t) = \frac{1}{\int_{\G_{LG}(t)} 1/\lambda^s \ \text{d}S(t)}\int_{\G_{LG}(t)}(\nabla_s \cdot \n)(\vel \cdot \n) \ \text{d}S(t).
\end{equation}
This formulation is useful numerically as it expresses $W(t)$ in terms of the bulk velocity $\vel$ without requiring the knowledge of the surface velocity $\svel$. An expression for the tangential component of the surface velocity can be derived, if required.

We now discuss the two critical events of encapsulation, after which a drop immediately enters a surface viscous regime, and liquid marble formation, when the interface becomes frozen. 

\subsubsection{Encapsulation}

\begin{figure}[h]
	\begin{center}
        	\includegraphics[width=\textwidth*19/20]{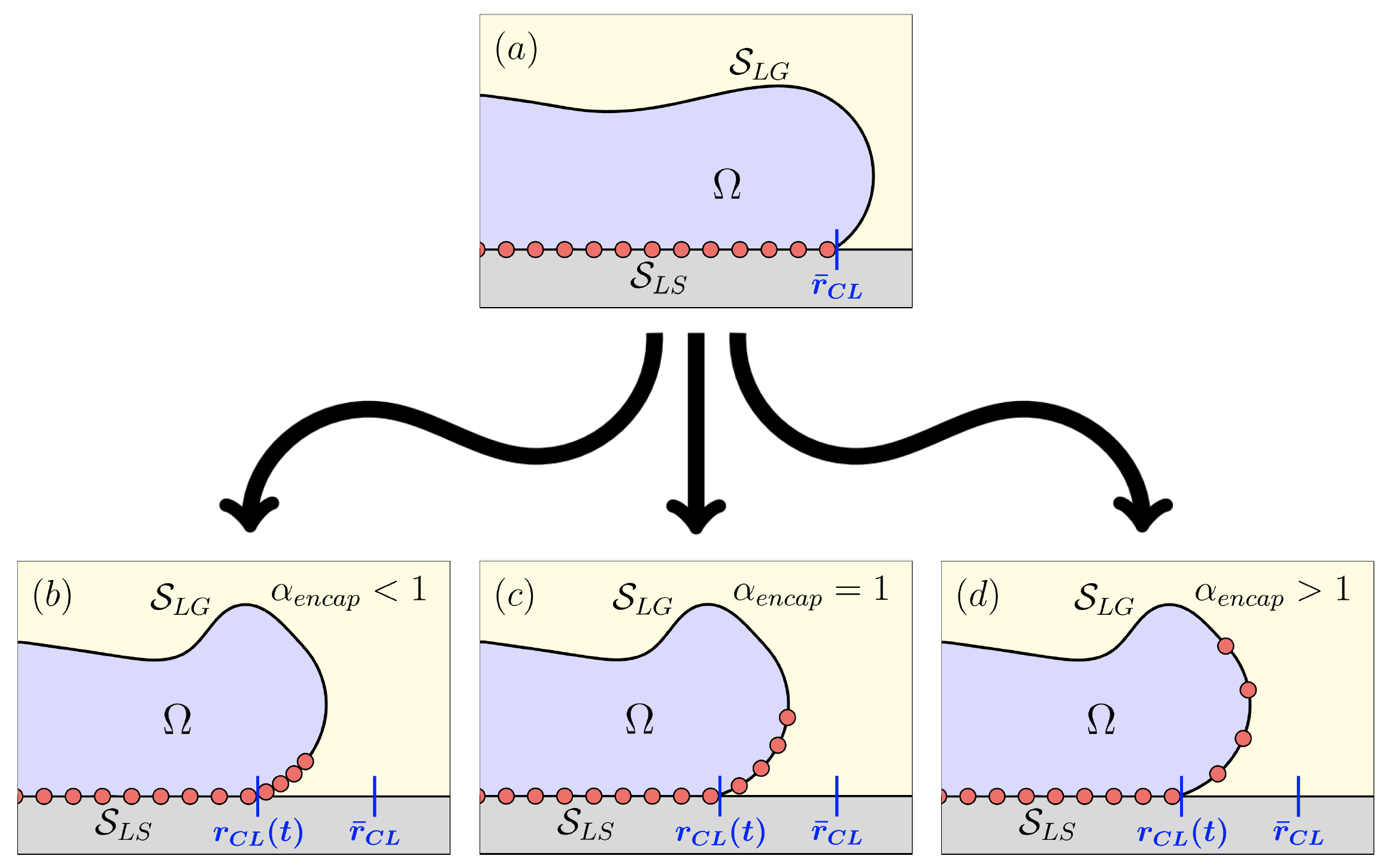} 	
	\end{center}
	    \caption{Diagram showing (a) the drop at maximum spread, with maximum contact line radius $\bar{r}_{CL}$ and a uniform coating of powder on its liquid-solid interface $\G_{LS}$, followed by retraction of the drop with the contact line reducing to $r_{CL}(t) < \bar{r}_{CL}$ at some time $t$. We have (b) $\alpha_{encap} < 1$, for more densely packed powder on the liquid-gas interface $\G_{LG}$, (c) $\alpha_{encap} = 1$, for the same packing on $\G_{LG}$, or (d) $\alpha_{encap} > 1$, for more sparsely packed powder on $\G_{LG}$.}
    \label{FIG:alpha_encap}
\end{figure}

In the pursuit of simplicity, we take the concentration on the liquid-solid interface $c_{LS}$, which corresponds to those particles which eventually end up on the drop's liquid-gas interface (i.e. not necessarily every particle the drop touches), to be constant and recall that in general the concentration on the powder-coated regions of the liquid-gas interface $c_{LG}$ can vary in time but not space. However, up until encapsulation we have found that in our model, the area of the powder coated region remains constant once the drop has rebounded and so $c_{LG}$ is also constant in time in this period. Consequently,  the total mass of adsorbed particles on $\G_{LS}$ is given by $M_{contact} = c_{LS} \cdot A_{contact}$, where the maximum contact area of the drop on the substrate is $A_{contact} = \pi \bar{r}_{CL}^2$ and $\bar{r}_{CL}$ denotes the maximum radius of the contact line.  Given that the total mass of adsorbed particles is conserved, at encapsulation we must have $M_{contact}= c_{LG} \cdot A_{encap}^{(0)} = c_{LS} \cdot A_{contact}$, where we recall $A_{encap}^{(0)}$ can include a contribution from a partially coated satellite drop as well as the encapsulated primary drop.  Consequently, we see that 
\begin{equation}{\label{EQN:alpha_enc}}
    \alpha_{encap} \equiv \frac{ A_{encap}^{(0)}}{A_{contact}}= \frac{c_{LS}}{c_{LG}},
\end{equation}
which gives a more physical picture of this parameter, which measures how the concentrations of powder differ on the two interfaces.  This raises interesting questions as to how the change in concentration occurs as particles on the liquid-solid interface become a part of the liquid-gas interface, but these are beyond the scope of the current work, where we simply assume this is instantaneous.  As shown in Figure \ref{FIG:alpha_encap}, there are three scenarios that can occur with regards to the powder coating as the contact line recedes from maximum spread depending on whether $\alpha_{encap}$ is larger, smaller, or equal to unity. 

Once encapsulated, the drop enters the surface viscous regime where $W(t) \not\equiv 0$, and surface area conservation for the powder region of $\G_{LG}$ no longer applies. An important consideration is that after a drop has been encapsulated, it is considered encapsulated for all future time (so for example, we assume that `holes' within the powder coating cannot re-open).

As will be discussed, in the surface viscous regime, so following drop encapsulation but prior to (potentially) deformed liquid marble formation, we must keep track of the surface area of the primary drop associated with the (spatially constant) powder concentration attaining its value at encapsulation ($c_{LG}$). This \emph{time dependent} surface area is called the `encapsulation area' $A_{encap}(t)$, and if the powder mass on the drop surface stays constant from maximum spread to encapsulation, then we just have $A_{encap}(t) = A^{(0)}_{encap}$, but this is more complex if satellite drop ejections remove powder. However, given that the concentration is always uniform in the powder coated region of the liquid-gas interface, $c$ is continuous through pinch-off events and in Appendix \ref{APP:TimeDep_Encap_Freeze} we show how this fact can be used to keep track of $A_{encap}(t)$ through satellite ejection events.

\subsubsection{Interfacial Freezing}\label{later}

Interfacial freezing is caused by the jamming of adsorbed particles on the drop surface which occurs at a critical concentration, denoted by $c_{freeze}$, which we assume is independent of $\text{We}$. Given that we have assumed the concentration of the particle coating is spatially constant post-encapsulation, particle concentration scales as the reciprocal of drop surface area, so the concentration reaching $c_{freeze}$ corresponds to the surface area reaching the `initial freezing area' $A_{freeze}^{(0)}$. Using the same arguments as for encapsulation, we can show that
\begin{equation}{\label{EQN:alpha_fre}}
    \alpha_{freeze} \equiv \frac{ A_{freeze}^{(0)}}{A_{contact}}=  \frac{c_{LS}}{c_{freeze}}.
\end{equation}
Similarly, we must keep track of the surface area of the primary drop associated with the powder attaining the critical freezing concentration ($c_{freeze}$), called the `freezing area' $A_{freeze}(t)$, as formulated in Appendix \ref{APP:TimeDep_Encap_Freeze}.

Notably, experiments showed that $\alpha_{freeze}$ does not depend on We, so $c_{LS}$ must also be independent of this parameter.  Therefore, in our simple model the dependence of $\alpha_{encap}$ on $\text{We}$ must indicate $c_{LG}=c_{LG}(\text{We})$.  This could be caused by higher impact speeds creating strong compressional surface dynamics that lower the effective value $c_{LG}$.  Another possibility is that $c_{LS}$ should depend on $\text{We}$, but these explanations are merely speculation and the topic clearly deserves further attention.

\subsubsection{Surface Viscous Regime}

The surface viscous regime lasts from drop encapsulation up to (potentially) liquid marble formation and in this period drop dynamics are affected by surface viscosity, which is captured via the effective surface tension (equations \ref{EQN:EffectiveSurfaceTension_Dimensional} and \ref{WEquation}). What remains is to provide the \emph{simplest} constitutive equation for the dilatational surface viscous coefficient $\lambda^s$ that satisfies the assumptions we have made in the construction of this model thus far. As such, we suppose $\lambda^s$ takes the following form:
\begin{equation}{\label{EQN:LambdaConstitutive}}
    \lambda^s(A, t) = \beta \cdot \frac{\left(A_{encap}(t) - A\right)}{\left(A - A_{freeze}(t)\right)},
\end{equation}
which holds for $A_{freeze}(t) < A \le A_{encap}(t)$, and where $\beta > 0$ is a dimensional parameter.  Notably, all the surface areas here refer to that of the primary drop only.

According to equation (\ref{EQN:LambdaConstitutive}), as $A$ approaches the freezing area $A_{freeze}(t)$, the surface viscous coefficient $\lambda^s$ diverges, so that surface viscous effects have a strong effect on drop dynamics. Surface viscous stresses provided by equation (\ref{EQN:LambdaConstitutive}) can delay the drop surface area from reaching $A_{freeze}(t)$ (as compared to a clean simulation) but do not prevent it from occurring indefinitely, and so when $A = A_{freeze}(t)$ we claim that interfacial freezing has occurred, and the drop is now a deformed liquid marble.

The surface viscous dynamics are most interesting and prominent when $A$ is close to the freezing area, so, for simplicity, we use the concentration $c_{LG}$ (and so a surface area of $A_{encap}(t)$) to signify negligible surface viscous effects in our constitutive relation.  If $A$ happens to increase back above $A_{encap}(t)$, we set $\lambda^s(A, t) = 0$ temporarily until $A \le A_{encap}(t)$ again.  

\subsubsection{Limits on the Effective Surface Tension}

Our constitutive equation ensures $\lambda^s\ge 0$, but no such condition exists on the curvature ($\nabla_s \cdot \n$) or normal velocity ($\vel \cdot \n$) at the interface $\G_{LG}$, hence the integral in the numerator on the right hand side of our effective surface tension 
\begin{equation}{\label{EQN:EffSurfTension_IntegralRepresentation}}
    \sigma_{\text{eff}}(A, t) = \sigma + \lambda^s(A, t) \left(\frac{\int_{\G_{LG}(t)}(\nabla_s \cdot \n)(\vel \cdot \n) \ \text{d}S(t)}{\int_{\G_{LG}(t)} 1 \ \text{d}S(t)}\right).
\end{equation}
 is able to take any real value. In particular, though rare, it is possible that $\sigma_{\text{eff}}$ can become negative for particularly exotic shapes near freezing coupled to  complex distributions of curvatures and normal velocities at its interface. In this case, small perturbations at the interface will grow very fast even over short intervals of time. Therefore, to prevent small numerical errors destroying the surface, we suppose that the minimum value for $\sigma_{\text{eff}}(A, t)$ is a small positive constant ($10^{-5}$ in simulations). 

\subsection{Viscous Bulk Equations}
We only expect viscous effects to be important during the impact itself (when there are thin films of liquid in contact with the substrate and viscosity is essential), with their influence minimal during rebound so that we have approximately inviscid bulk dynamics. For this reason, we exploit the simple effective surface tension model developed earlier for inviscid bulk flow and apply it to a model with viscous bulk flow. What this means is that we take the conventional boundary condition for the normal stress at the interface between two viscous flows, and replace the surface tension with our \emph{effective} surface tension, that is
\begin{equation}{\label{EQN:NormalStress_Viscous}}
    \n \cdot \left(\stress_G - \stress\right) \cdot \n = \sigma_{\text{eff}}(A, t)\left(\nabla_s \cdot \n\right) \quad \text{on} \ \G_{LG}.
\end{equation}
where the effective surface tension $\sigma_{\text{eff}}(A, t)$ is given by equation (\ref{EQN:EffSurfTension_IntegralRepresentation}) with the constitutive equation (\ref{EQN:LambdaConstitutive}) for $\lambda^s(A, t)$.

For the tangential stress balance equation, we continue with the conventional boundary condition for free surface flows, namely:
\begin{equation}{\label{EQN:TangentialStress_Viscous}}
    \n \cdot \left(\stress_G - \stress\right) \cdot \tang = 0.
\end{equation}

The constitutive equations for the liquid and gas bulk stress tensors in this viscous formulation are given by, respectively,
\begin{equation}{\label{EQN:Bulk_Liquid_StressTensor}}
    \stress = -p\identity + \mu\left(\nabla \vel + \left(\nabla \vel\right)^T\right), \qquad \stress_G = -p_G\identity + \mu_G\left(\nabla \vel_G + \left(\nabla \vel_G\right)^T\right).
\end{equation}
The gas dynamics in principle could be neglected, but it is convenient to include them as they are retained in the volume-of-fluid method. The conservation of momentum equations within the liquid and gas then take the form of the incompressible Navier-Stokes equations, that is,
\begin{equation}{\label{EQN:NavierStokes}}
     \nabla \cdot \vel = 0,\qquad   \rho\left[ \frac{\partial \vel}{\partial t} + (\vel \cdot \nabla)\vel \right] = \nabla\cdot\stress  + \rho\grav  \quad \text{in} \ \Omega,
\end{equation}
in the liquid drop, and
\begin{equation}{\label{EQN:NavierStokes_Gas}}
 \nabla \cdot \vel_G = 0,\qquad    \rho_G\left[ \frac{\partial \vel_G}{\partial t} + (\vel_G \cdot \nabla)\vel_G \right] = \nabla\cdot\stress_G + \rho_G\grav \quad \text{in} \ \Omega_G,
\end{equation}
in the exterior gas. 

Finally, we impose the following symmetry boundary conditions on the bulk flows within the liquid drop and exterior gas at the axis of symmetry:
\begin{equation}{\label{EQN:Bulk_Viscous_BCs_at_Axis}}
    \frac{\partial}{\partial r}\left(\vel \cdot \e_z\right) = \frac{\partial}{\partial r}\left(\vel_G \cdot \e_z\right) =    \vel \cdot \e_r = \vel_G \cdot \e_r=0 \quad \text{at} \ r = 0.
\end{equation}

\section{Setup for the Computational Model}{\label{SEC:Simulations_Model_Refinement}}

We use popular open-source software Basilisk \cite{Basilisk} which has been used to, for example, solve the Serre–Green–Naghdi equations \cite{Popinet2015quadtree}, simulate the turbulent regime in Rayleigh-B\'enard convection cells \cite{Castillo2016reversal, Castillo2017turbulent}, and to simulate complex bubble dynamics in the presence of surface tension \cite{Fuster2018all}. The numerical methods implemented in Basilisk have been validated for simulations of multiple problems relating to multiphase flows with moving interfaces \cite{Fuster2009numerical}, with many examples available for free on the Basilisk website \cite{Basilisk}. Notably, the volume-of-fluid solver in Basilisk is capable of naturally handling changes in topology which are crucial to the liquid marble process, and can solve the Navier-Stokes equations. As far as we are aware, this is the first attempt to incorporate a dilatational surface viscosity into this volume-of-fluid method using Basilisk. A straightforward incorporation of the effective surface tension led to Basilisk quickly becoming unstable for large drop deformations, most likely due to the combination of explicit time stepping with the need to accurately resolve interfacial quantities (including surface integrals), so we had to devise our own implicit-in-time iterative process for the effective surface tension's value \cite{pritchard_phd}.  This method worked remarkably well, both in terms of stability and as verified by excellent agreement with boundary element method simulations for suspended inviscid drops undergoing large-amplitude oscillations with (constant) $\lambda^s$ varying across multiple orders of magnitude, as considered in \cite{pritchard_2021}.

Before presenting our results, we note that the experimental liquid marble formation by drop impact is a sensitive process and can exhibit large variability in shapes and drop behaviour, as will be seen for results at the moment of encapsulation and for the liquid marbles themselves.  Therefore, perfect agreement of simulations to experiments should not be expected.

\subsection{Choice of $\beta$}
\begin{figure}
	\begin{center}
        	\includegraphics[width=\textwidth*19/20]{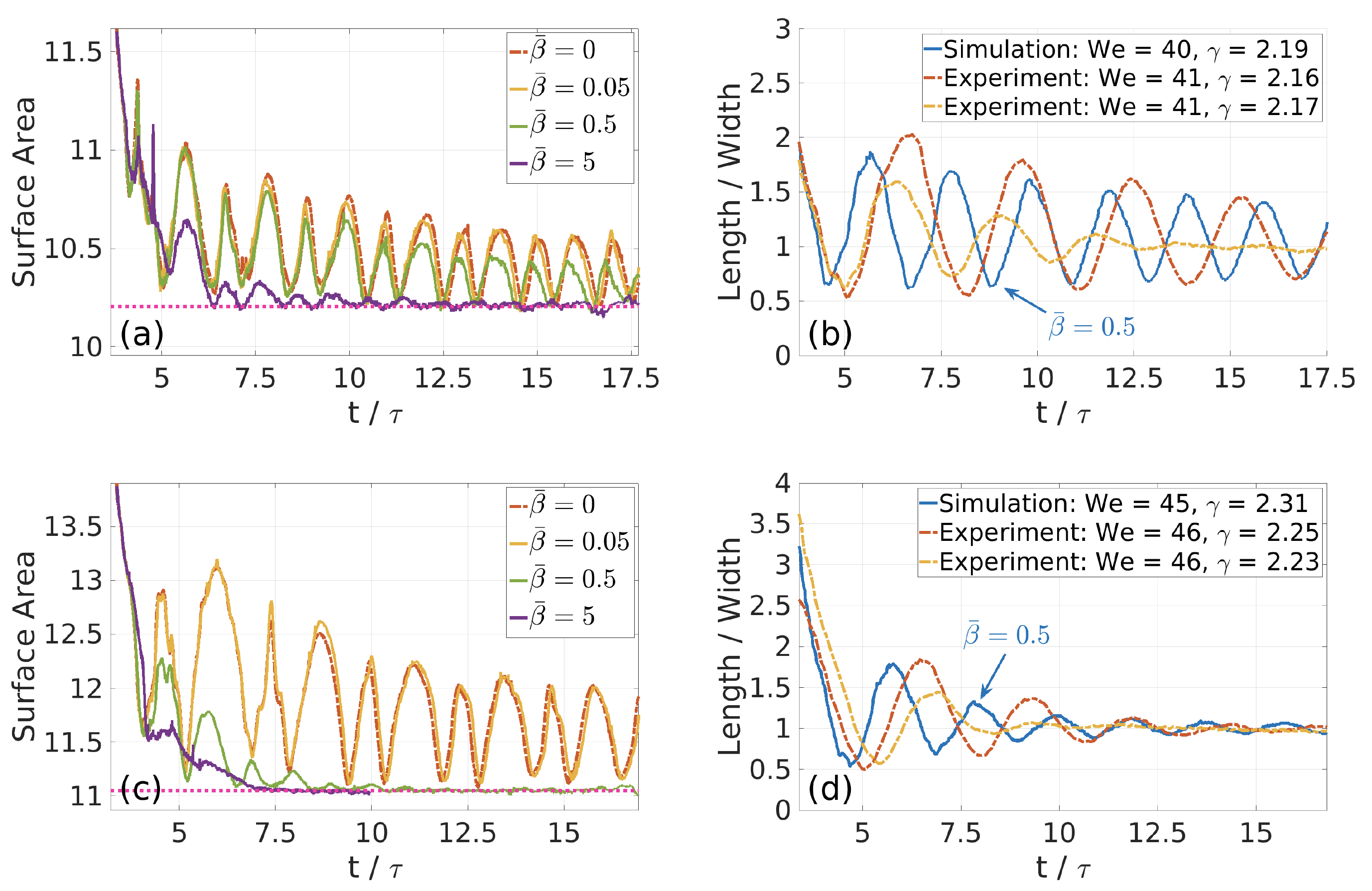} 	
	\end{center}
  \caption{Plots from the moment of encapsulation to when the drop reaches the apex of its rebound from the substrate. Simulations are shown for two different We numbers (a-b) and (c-d) with the decay of surface area for different $\bar{\beta} = 0.05, 0.5, 5$ shown in (a) and (c) and a comparison of the aspect ratio of the drops to experiments, for the chosen value of $\bar{\beta}=0.5$, shown in (b) and (d). Pink dotted lines in (a) and (c) show the surface area of the sphere, the minimum surface energy state, which differs between the two plots because of prior satellite droplet ejections. Encapsulation in experiments is matched to occur at the start of the plot. Surface area and time are given in dimensionless units, using a length-scale of $R_0$ and capillary time-scale of $\tau = \sqrt{\rho R_0^3 / \sigma}$.}
    \label{FIG:beta_comparison_We=40_45}
\end{figure}

{\label{SEC:ChoosingBeta}}
The constitutive equation (\ref{EQN:LambdaConstitutive}) for the dilatational surface viscous coefficient $\lambda^s$  contains a coefficient $\beta > 0$ that we choose to best match our simulations for liquid marble formation with experiments. To do so, we define a dimensionless parameter $\bar{\beta}=\beta/\sqrt{\sigma \rho R_0^3}$. To find $\bar{\beta}$, we consider the range of We in which spherical liquid marbles are formed, namely between We = 40 and We = 45, as it is in this regime that the surface viscous model has a key role in damping drop oscillations. In Figure \ref{FIG:beta_comparison_We=40_45}(a) we see the effects on the drop surface area from varying $\bar{\beta}$ across orders of magnitude for the We = 40 simulation, noting that $\bar{\beta} = 0.05$ provides negligible damping with respective to the clean case, that $\bar{\beta} = 0.5$ provides \emph{some} damping, and that $\bar{\beta} = 5$ causes a quick reduction to the surface area of the sphere (indicating likely spherical marble formation). In Figure \ref{FIG:beta_comparison_We=40_45}(b) we see the ratio of the length of the drop's primary and secondary axes post-encapsulation for the We = 40 and $\bar{\beta} = 0.5$ simulation, showing that it is within the extremes observed in experiments with We = 41. In Figure \ref{FIG:beta_comparison_We=40_45}(c) we then see that for the We = 45 simulation, $\bar{\beta} = 0.5$ shows a quick reduction to the spherical surface area, with the decay in drop oscillations similarly shown to be within experimental observations for We = 46 in Figure \ref{FIG:beta_comparison_We=40_45}(d). In summary, while $\bar{\beta} = 0.05$ and $\bar{\beta} = 5$ result in spherical liquid marble formation in neither and both of the two cases, respectively, $\bar{\beta} = 0.5$ provides non-negligible damping for We = 40, and spherical marble formation in We = 45, as desired.

Of note in Figure \ref{FIG:beta_comparison_We=40_45}(d) is that there are still visible oscillations by the end of our simulation. For a drop undergoing oscillations with an inviscid bulk flow and dilatational surface viscous effects given by our effective surface tension (\ref{EQN:EffSurfTension_IntegralRepresentation}), it is known that for $\lambda^s > 0$ the rate of decay of the oscillations due to dilatational surface viscosity vanishes as the amplitude goes to zero, see \cite{pritchard_2021}. In Figure \ref{FIG:beta_comparison_We=40_45}(d) we appear to see that damping effects have diminished significantly soon after $t = 10$, with only bulk viscosity now acting to decay oscillations. However, these oscillations are sufficiently small as to claim a spherical liquid marble is formed.

\section{Results}{\label{SEC:Simulation_Results}}

\begin{figure}
 	\begin{center}
        	\includegraphics[width=\textwidth*19/20]{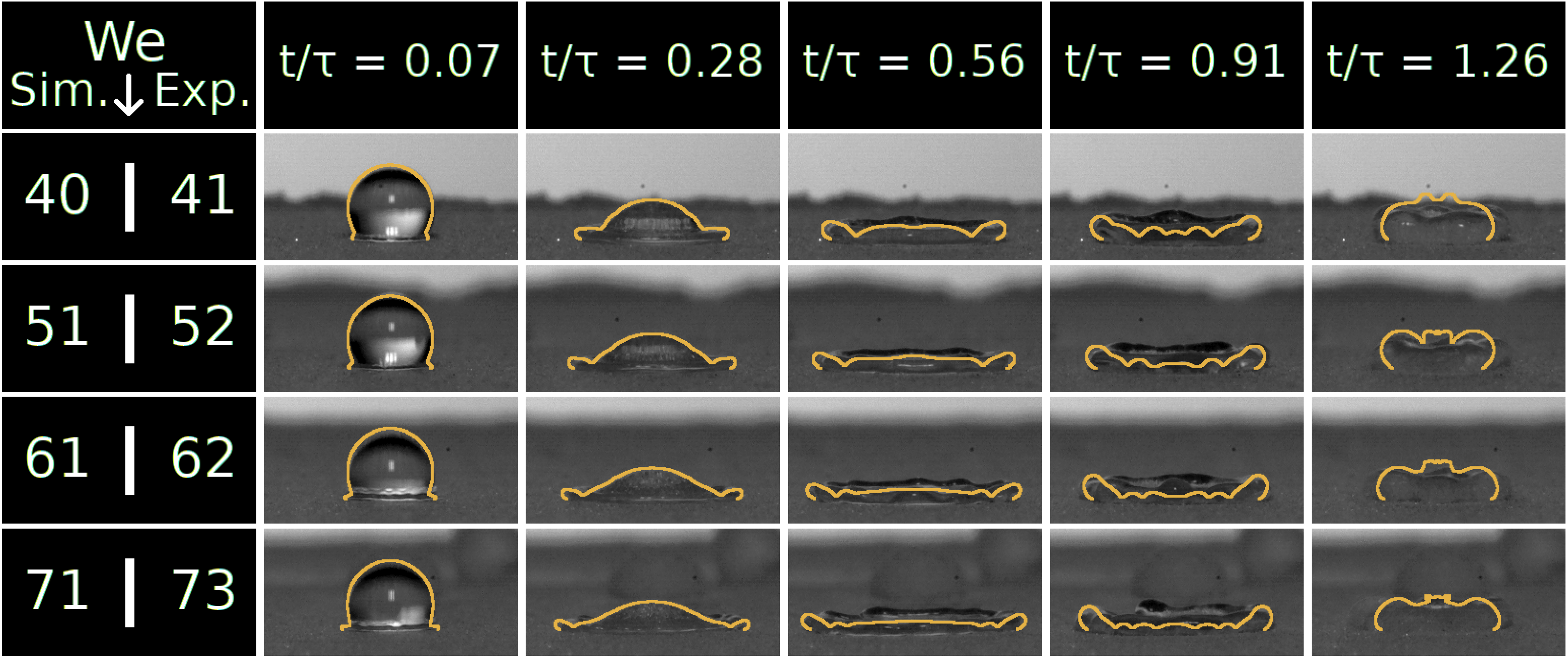} 	
	\end{center}
    \caption{Comparison of drop profiles for simulations (yellow line) overlaid onto images taken from our experiments at We $\approx 40, 51, 61, 71$, at multiple instances in (dimensionless) time. Snapshots are provided shortly after contact is made with the powder bed, leading up to and passing through maximum spread, also during retraction, and immediately prior to the formation of a vertical jet which leads to the drop rebounding from the powder bed.}
    \label{FIG:DropletImpact_Shapes}
\end{figure}

For the spreading and retraction phases, in which there is a liquid-solid contact, excellent agreement is obtained between simulations and experiments for both the shapes, maximum spread, and contact times with the powder bed (Figure~\ref{FIG:DropletImpact_Shapes}).  

\subsection{Overview of Liquid Marble Formation}{\label{SEC:LiquidMarble_SimulationOverview}}

\begin{figure}
	\begin{center}
        	\includegraphics[width=\textwidth*19/20]{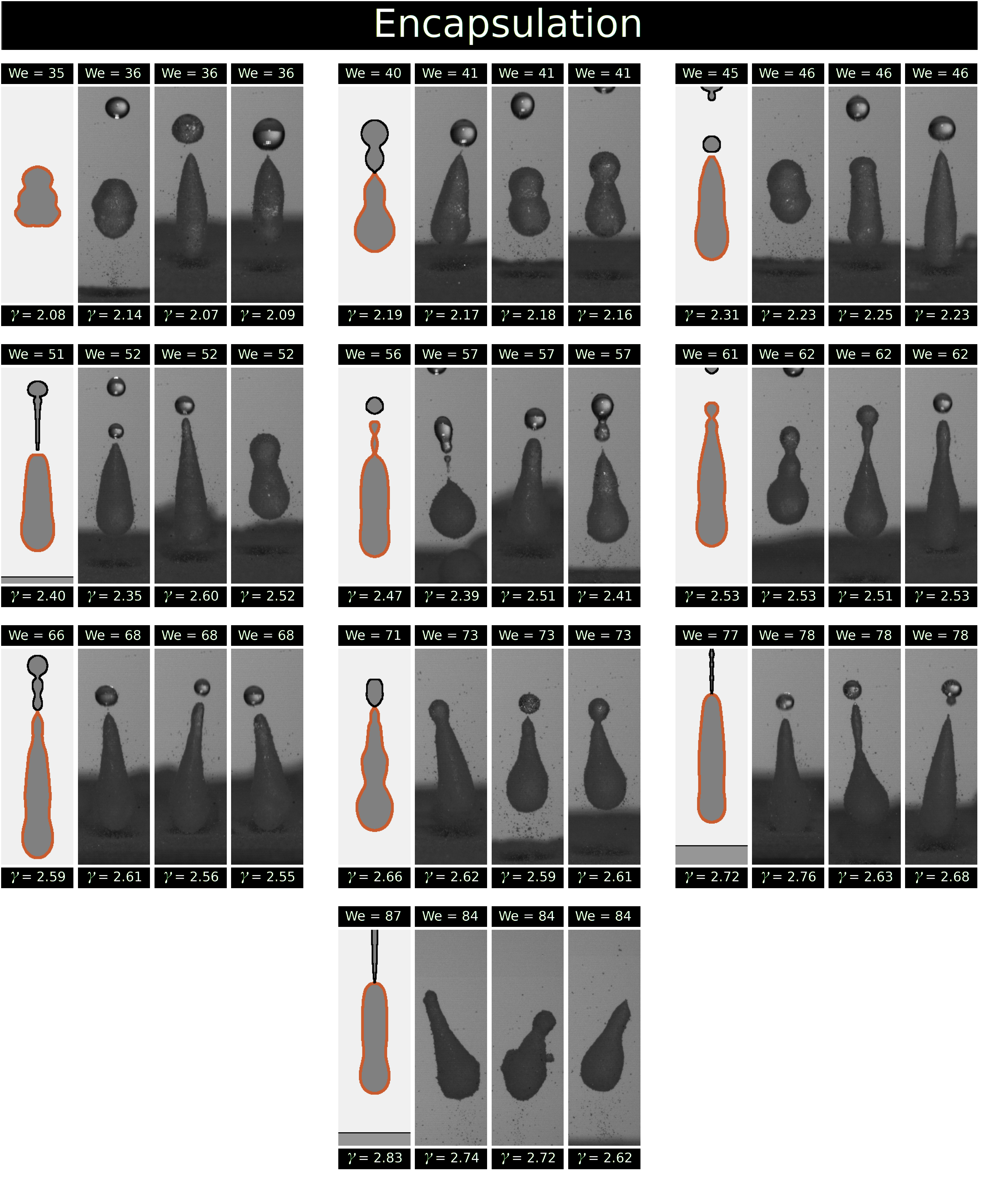} 	
	\end{center}
    \caption{Drop profiles at encapsulation for $35 \le$ We $\le 87$. The red outlines indicate the primary drop at the moment of encapsulation - any other drops in the simulation are ignored in the subsequent dynamics. The impact Weber number for each simulation or experiment is provided above each drop profile, with the spreading factor provided below.}
    \label{FIG:LiquidMarbles_Encap}
\end{figure}

The drop shapes at encapsulation for our simulations at 10 different We are provided in Figure \ref{FIG:LiquidMarbles_Encap} alongside images from experiments for (almost) the same impact Weber number. In general, the qualitative agreement is good, with the drop shape at encapsulation becoming more elongated for higher We, owing to the fact that more powder has been adsorbed from the substrate. For the We = 87 simulation and accompanying experiments, encapsulation and freezing occur simultaneously.

The agreement is most clearly seen in the movies (Supplemental Material \cite{SM}), which are provided for $\text{We}=45$ (spherical liquid marble), $\text{We}=51$ (deformed liquid marble) and $\text{We}=71$ (more deformed liquid marble) where points of encapsulation and freezing are highlighted for both experiments and simulations. Note that in the simulations satellite drops are removed from the computation once they detach from the primary drop and computations finish when the drop is frozen (and thus the drop appears stationary).

Figure \ref{FIG:LiquidMarbles_Encap} shows that for a given impact Weber number, there is significant experimental variation in drop shapes at the moment of encapsulation, so it is unsurprising that we do not see \emph{perfect} agreement in our simulations, although we consider the resemblance to be good overall. Interestingly, a simulation for a particular We can resemble experiments that have a somewhat \emph{different} We. A particular example of this is the simulation for We = 45 and the experiment with We = 52. Later, for the same simulation, we see that the initial dynamics following encapsulation match well with an experiment with We = 57. A likely explanation for this is loss of energy in deforming the powder bed, which is not currently accounted for in the simulations.

\begin{figure}
  	\begin{center}
        	\includegraphics[width=\textwidth*19/20]{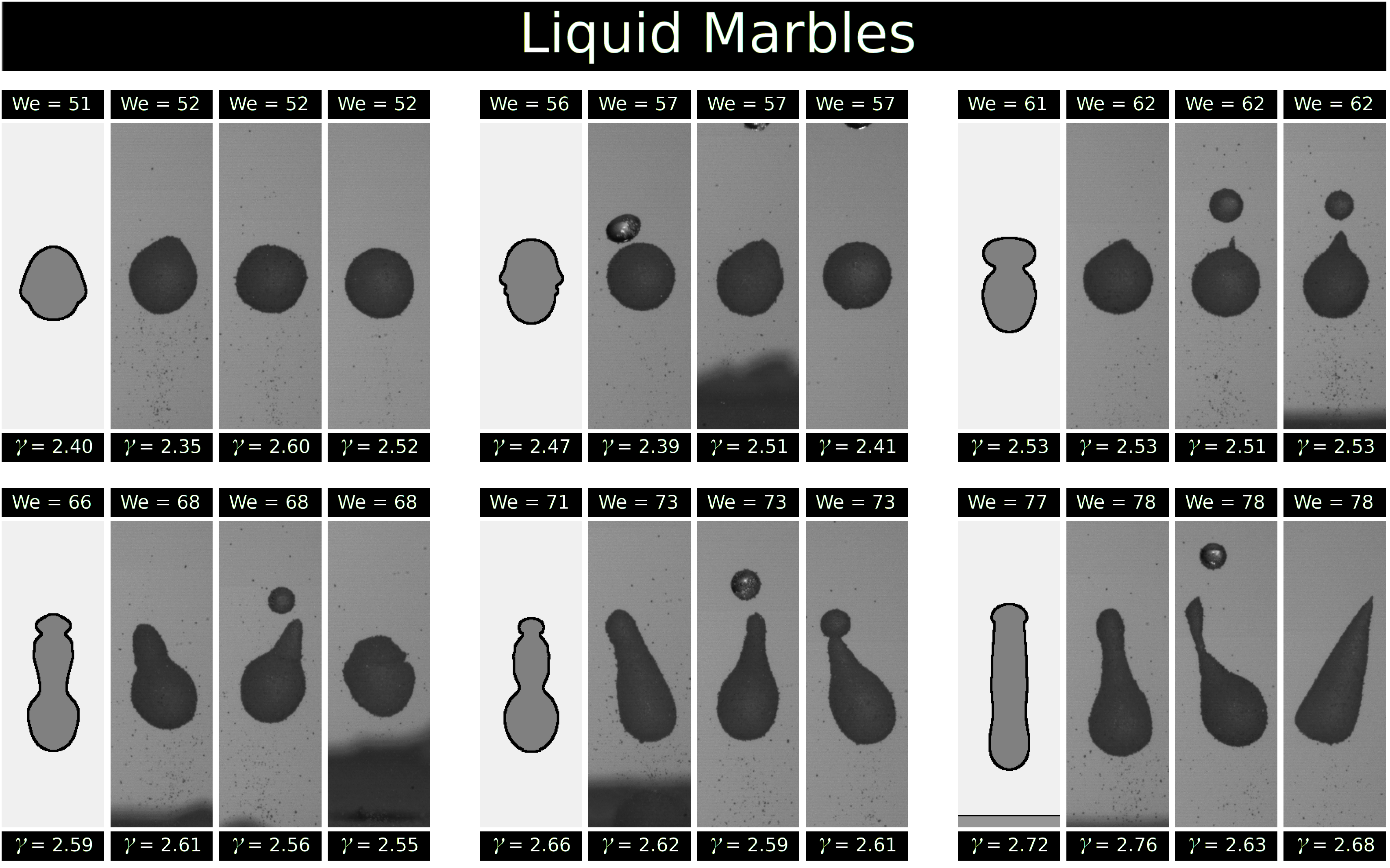} 	
	\end{center}    \caption{Drop profiles for deformed liquid marbles obtained for $51 \le$ We $\le 78$. The impact Weber number for each simulation or experiment is provided above each drop profile, with the spreading factor provided below.}
    \label{FIG:LiquidMarbles_DeformedMarbles}
\end{figure}

In Figure \ref{FIG:LiquidMarbles_DeformedMarbles}, we see all deformed liquid marbles created in our simulations with images from experiments of all the drops initially shown at encapsulation in Figure \ref{FIG:LiquidMarbles_Encap}. The low impact Weber number simulations are absent as they either do \emph{not} form liquid marbles (We = 35, 40) or only form spherical marbles (We = 45), which is visually uninteresting. Also not included is the We = 87 case, as the liquid marble is formed at encapsulation and is shown in Figure \ref{FIG:LiquidMarbles_Encap}. We see in simulations and experiments that the increase in We (and in general the spreading factor) moves us from obtaining almost spherical liquid marbles to (sometimes intricate, see for example the experimental shape for We = 78 and $\gamma = 2.63$) elongated liquid marbles. Again we highlight that the shapes of the liquid marbles in experiments can vary considerably even when the impact Weber numbers and spreading factors are similar, so given this, the resemblance of our simulations to the shapes seen in experiments is good.

\subsection{Analysis of the Regimes of Liquid Marble Formation}
We will now discuss the different outcomes that emerge from these simulations; no liquid marble formation, spherical liquid marble formation, and deformed liquid marble formation.

\subsubsection{No Liquid Marble Formation}

We have chosen the value of $\beta$ in the constitutive equation for $\lambda^s$ such that simulations with We $<$ 45 do \emph{not} form liquid marbles of any kind. As we should therefore expect, the surface area of the drop for We = 40 does not differ greatly from the clean case, and the effective surface tension does not take values far from unity. 

\subsubsection{Spherical Liquid Marble Formation}{\label{SEC:Spherical_LM_Formation}}

\begin{figure}
	\begin{center}
        	\includegraphics[width=\textwidth*19/20]{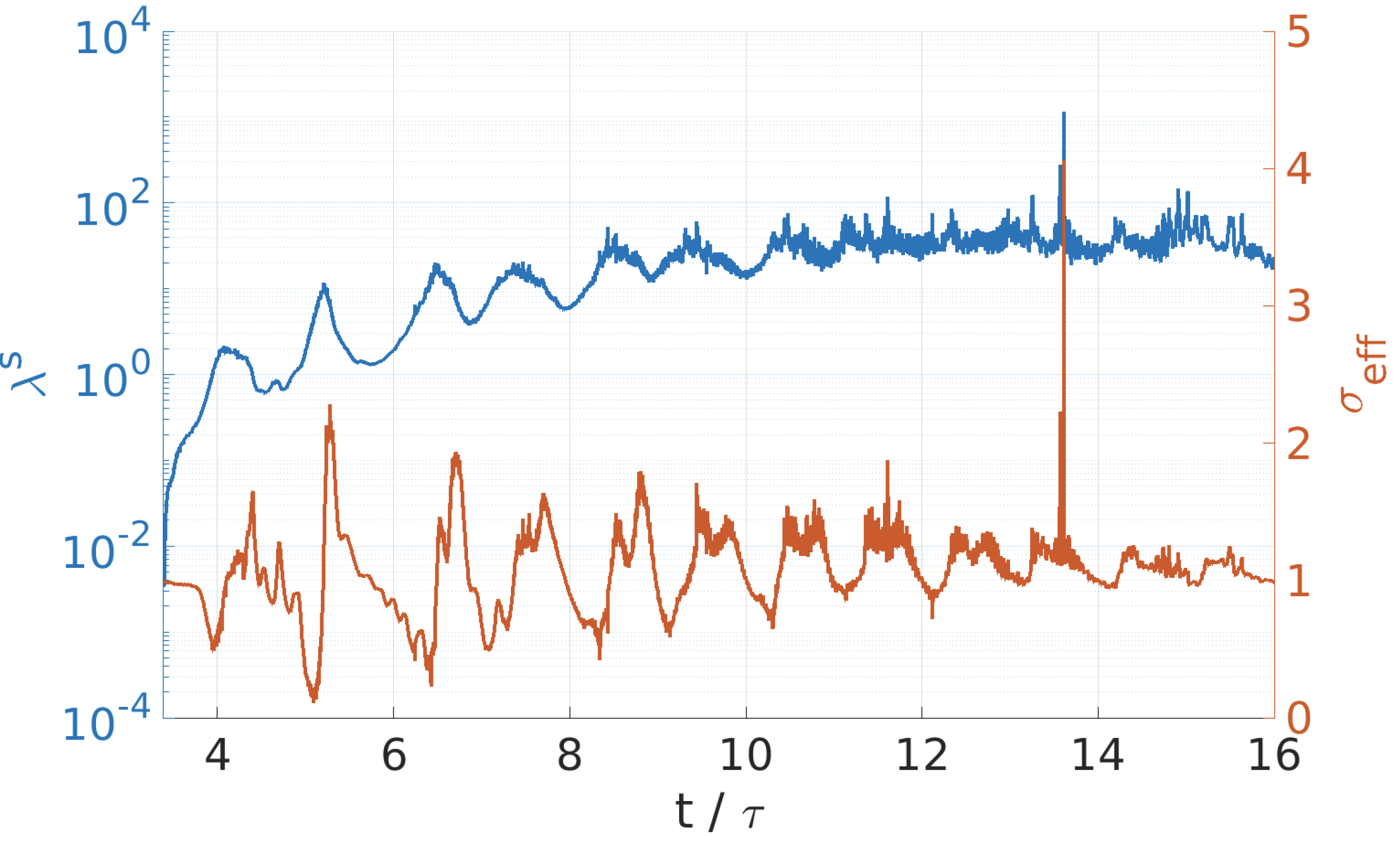} 	
	\end{center}
    \caption{The dilatational surface viscous coefficient $\lambda^s$ (log-scale) and effective surface tension $\sigma_{\text{eff}}$ over the course of the We = 45 simulation.}
    \label{FIG:LM_We=45}
\end{figure}

Recall that a spherical liquid marble is formed for We = 45 by our choice of $\beta$. Provided in Figure \ref{FIG:LM_We=45} is the value of $\lambda^s$ (in log-scale) and the effective surface tension $\sigma_{\text{eff}}$ over the course of the simulation. We note that as expected, as the surface area approaches the sphere (which is \emph{very} close to the freezing area in this case), $\lambda^s$ takes on large values (on the order of 10). The effective surface tension \emph{does} experience large changes as it nears zero and briefly exceeds $\sigma_{\text{eff}} = 2$ in the initial stages of the post-encapsulation dynamics, but as $\lambda^s$ becomes larger, $\sigma_{\text{eff}}$ moves to oscillate around unity, owing to the greatly reduced drop velocities due to the surface viscous damping effects. We see a brief spike in both the surface viscous coefficient $\lambda^s$ and therefore also in the effective surface tension $\sigma_{\text{eff}}$ prior to $t = 14$; this is due to the surface area of the drop being very close to the freezing area by this point in the simulation and so slight variations in the surface area approximation are magnified in the calculation of $\lambda^s$.

\begin{figure}[H]
	\begin{center}
        	\includegraphics[width=\textwidth*19/20]{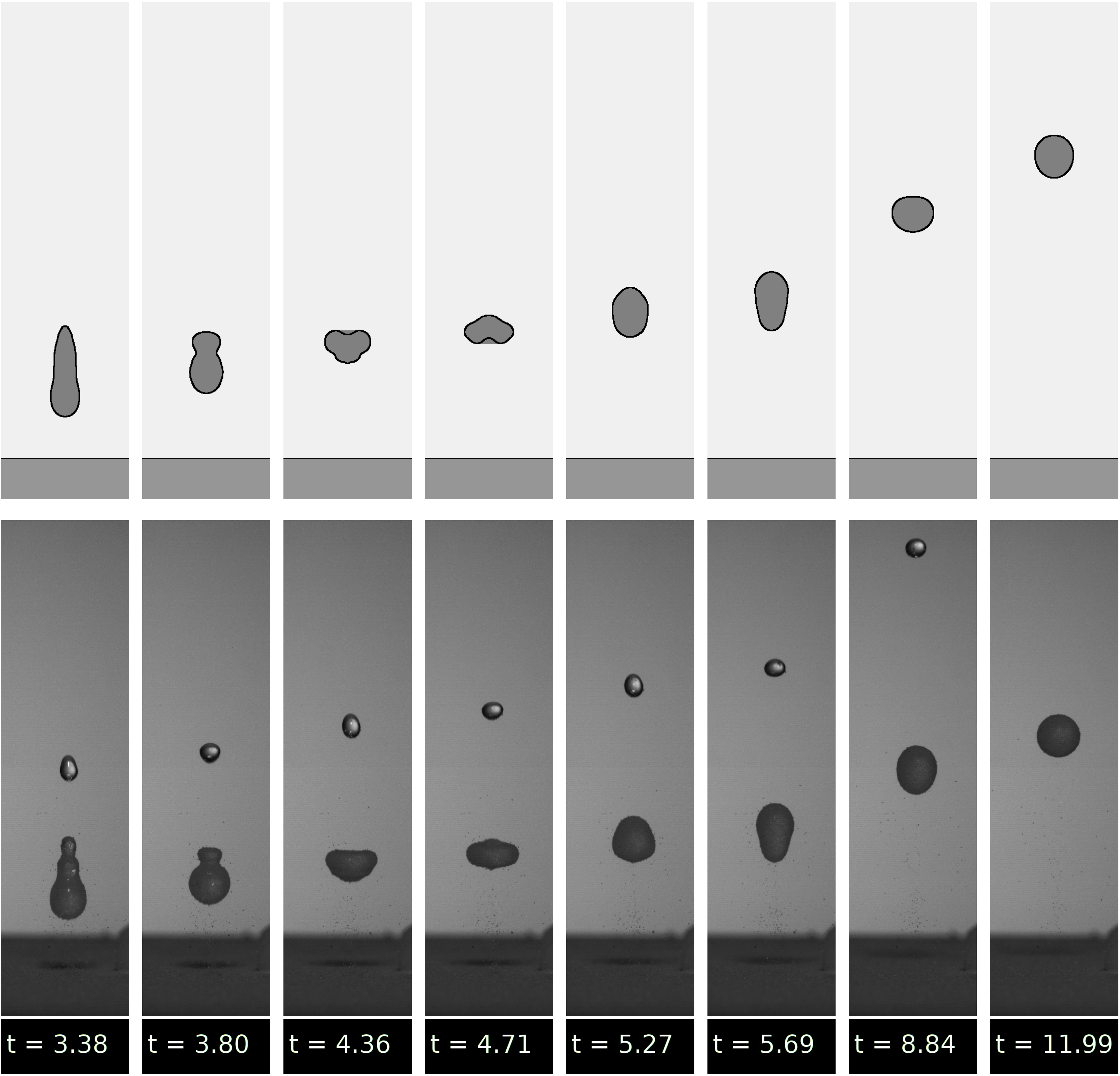} 	
	\end{center}
    \caption{Drop profile comparisons from the moment of encapsulation between the simulation with We = 45 and $\gamma = 2.31$, and a powder bed experiment with We = 57 and $\gamma = 2.45$. The dimensionless time taken between images in the experiment is the same as that in the simulation, and $t = 3.38$ is the time of encapsulation for the primary drop in the simulation.}
    \label{FIG:LiquidMarbles_We=45_Comparison}
\end{figure}

As we have seen in Figure \ref{FIG:beta_comparison_We=40_45}, the decay rate of oscillations in the We = 45 simulation sits between the extremes observed in experiments. However, it is not a We = 46 experiment with which this simulation bears the greatest resemblance; as alluded to earlier, there are examples of simulations and experiments showing good agreement (at least for a short time) even though they have different impact Weber numbers. An example of this is shown here in Figure \ref{FIG:LiquidMarbles_We=45_Comparison} comparing drop profiles from the moment of encapsulation to the formation of a spherical marble, between the simulation with We = 45 and $\gamma = 2.31$ and an experiment with We = 57 and $\gamma = 2.45$. The first six pairs of images from this figure highlight the initial window of time in which the two show very good agreement, and the final two pairs taken much later, after the simulation and experiment have drifted out of phase with one another. 

\subsubsection{Deformed Liquid Marble Formation}
In this subsection we consider characteristic cases leading to slightly deformed marbles and more elongated profiles.

\begin{figure}[H]
	\begin{center}
        	\includegraphics[width=\textwidth*19/20]{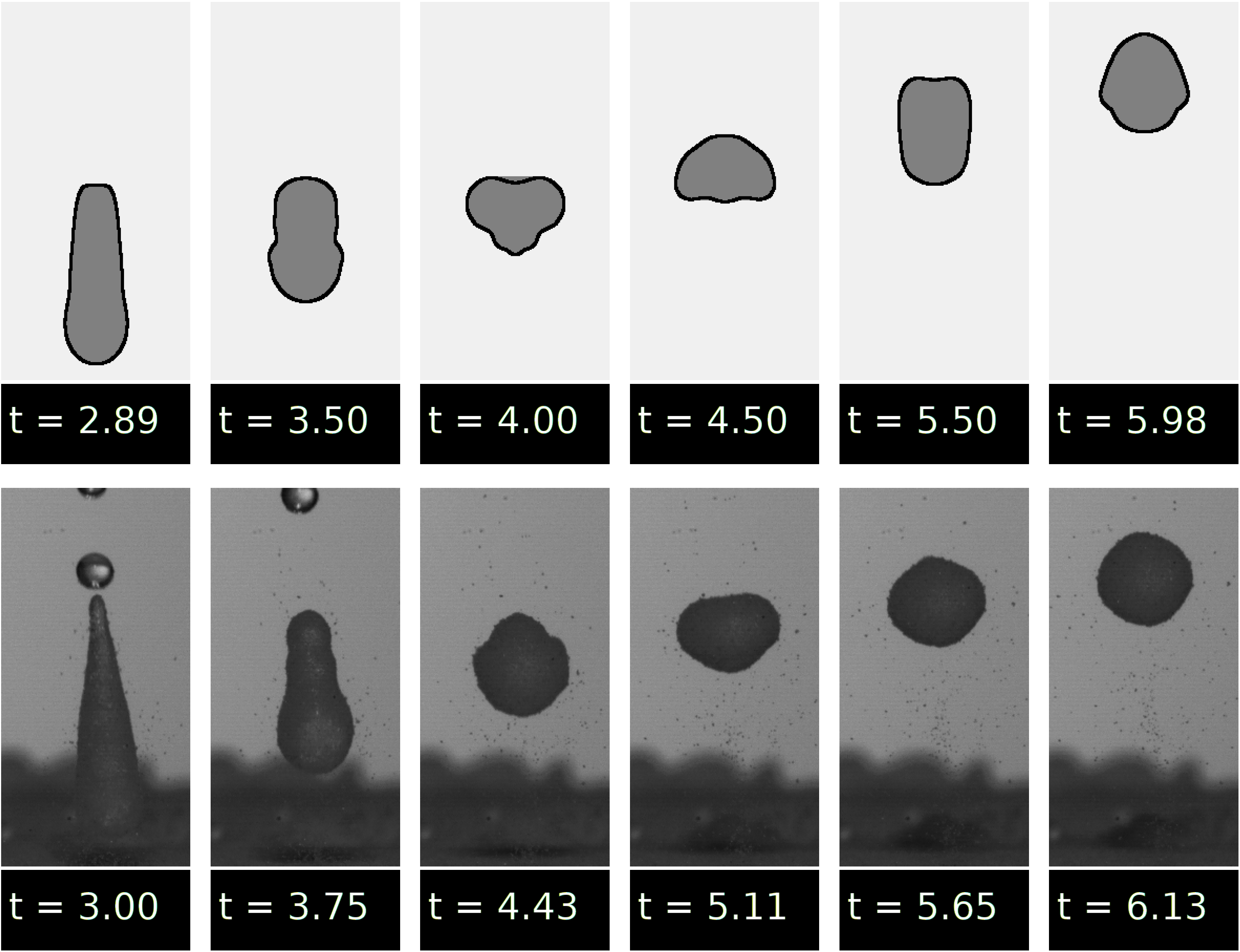} 	
	\end{center}	
	\caption{Drop profile comparison from the moment of encapsulation to the moment of deformed liquid marble formation between the simulation with We = 51 and $\gamma = 2.40$, and a powder bed experiment with We = 52 and $\gamma = 2.60$. The dimensionless time $t = 0.00$ corresponds to the moment the drop first comes into contact with the substrate. }
    \label{FIG:LiquidMarbles_We=51_Dynamics}
\end{figure}

\begin{flushleft}
\emph{We=51}
\end{flushleft}
The We = 51 simulation is the first example of a deformed liquid marble forming from our simulations. In Figure \ref{FIG:LiquidMarbles_We=51_Dynamics} we see the drop shape evolution from the moment of encapsulation to the moment of slightly deformed spherical marble formation for the We = 51 ($\gamma = 2.40$) simulation and a powder bed experiment with We = 52 ($\gamma = 2.60$). There are similarities between these results; liquid marble formation occurs approximately 3 units of (dimensionless) time after encapsulation in both cases (and surface viscous dynamics `kick in'), and the encapsulated shape is much more elongated than the eventual marble. Further, the encapsulated drop is able to substantially change its shape (becoming short and wide in the fourth panel of both cases), and the vertical distance travelled is very similar. Collectively, we take this as evidence of good qualitative agreement between the simulation based on our surface viscous model and experimental data.

A point of difference between the simulation and experiment in Figure \ref{FIG:LiquidMarbles_We=51_Dynamics} is that the bottom of the drop in the experiment appears more rigid and unable to deform as easily as the simulation. This is likely due to powder being more tightly packed on the bottom of the drop, which in turn is related to it being the last region to leave the substrate; our model by contrast assumes a spatially constant powder concentration on the liquid-gas interface and so neglects any such effect.

\begin{flushleft}
\emph{We=61}
\end{flushleft}
For higher impact Weber number cases, where (see Figure \ref{FIG:LM_Radius_Alpha}) $\alpha_{encap}$ is moving closer to $\alpha_{freeze}$ (so $A_{encap}^{(0)}$ is moving closer to $A_{freeze}^{(0)}$), leaving only a small window of time between encapsulation and freezing, and so causing the deformed liquid marbles to be less spherical and more elongated (as their corresponding shapes at encapsulation are) when they form. For We $= 61$ we obtain such an elongated deformed liquid marble within our simulation, and now describe three interesting phenomena also exhibited, which are likewise observed in experiments.

\begin{figure}
	\begin{center}
        	\includegraphics[width=\textwidth*19/20]{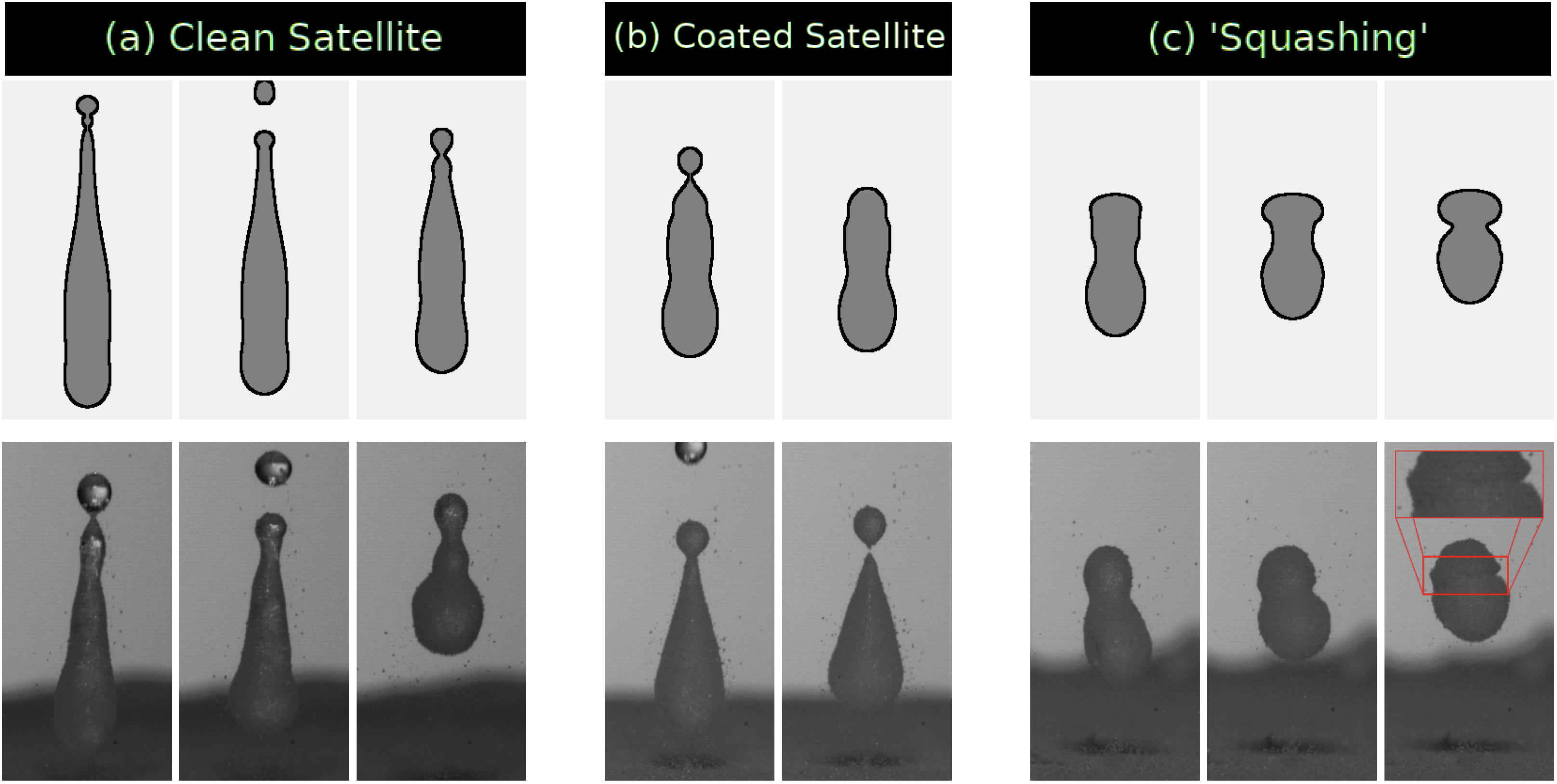} 	
	\end{center}
    \caption{Comparisons of interesting phenomena observed in the We = 61 ($\gamma$ = 2.53) simulation, with examples of the same phenomena occurring in our drop impact experiments: (a) We = 62 ($\gamma = 2.53$) experiment showing ejection of a clean satellite followed by a continuous reduction in surface area to encapsulation in the right-most image. (b) We = 62 ($\gamma = 2.53$) experiment showing ejection of a satellite post-encapsulation. (c) We = 68 ($\gamma = 2.56$) experiment showing the `squashing' of the drop surface leading to deformed liquid marble formation in the right-most image.}
    \label{FIG:We=61_phenomena}
\end{figure}

Firstly, this liquid marble simulation (We $= 61, \gamma = 2.53$) shows that encapsulation does \emph{not} always occur immediately following a satellite drop ejection. There are many examples in the experiments where, like this simulation, encapsulation is due to a continuous reduction in surface area, as the drop attempts to minimise its surface energy, rather than an ejection event. In Figure \ref{FIG:We=61_phenomena}(a) we see a very similar encapsulation scenario occurring in a We = 62 ($\gamma = 2.53$) experiment, whereby a \emph{clean} satellite drop is ejected; a short period of time passes in which the drop retracts in on itself, and \emph{then} encapsulation occurs.

Also seen in this simulation is the ejection of a \emph{fully}-coated satellite drop which leaves the primary drop with a concentration of powder \emph{greater} than at encapsulation (due to the powder coated surface area shrinking post-encapsulation). Shown in Figure \ref{FIG:We=61_phenomena}(b) is a similar post-encapsulation satellite drop ejection occurring in an experiment with We = 62 and $\gamma = 2.53$ (recall that satellite drops are removed from the computational domain in liquid marble simulations).

The moments leading up to the creation of the deformed liquid marble for We = 61 simulation is shown in Figure \ref{FIG:We=61_phenomena}(c).  Due to the small (reduced) value of the effective surface tension, a region of relatively high (negative) curvature is permitted to form and is sustained, so that the upward motion of the drop following rebound compresses the shape and squashes this negative curvature region until the freezing area is reached. This `squashing' of the drop shape is not observed in our simulations without surface viscosity, nor in our drop impact experiments onto a rigid superhydrophobic substrate, because the base surface tension of water is sufficiently strong to smooth these regions before this can occur. There is an example of this occurring in a powder bed experiment and is shown alongside the simulation in Figure \ref{FIG:We=61_phenomena}(c) with We = 68 and $\gamma = 2.56$, with a crevice forming in the experimental liquid marble similar to what is shown in our simulation.

\begin{figure}[H]
	\begin{center}
        	\includegraphics[width=\textwidth*19/20]{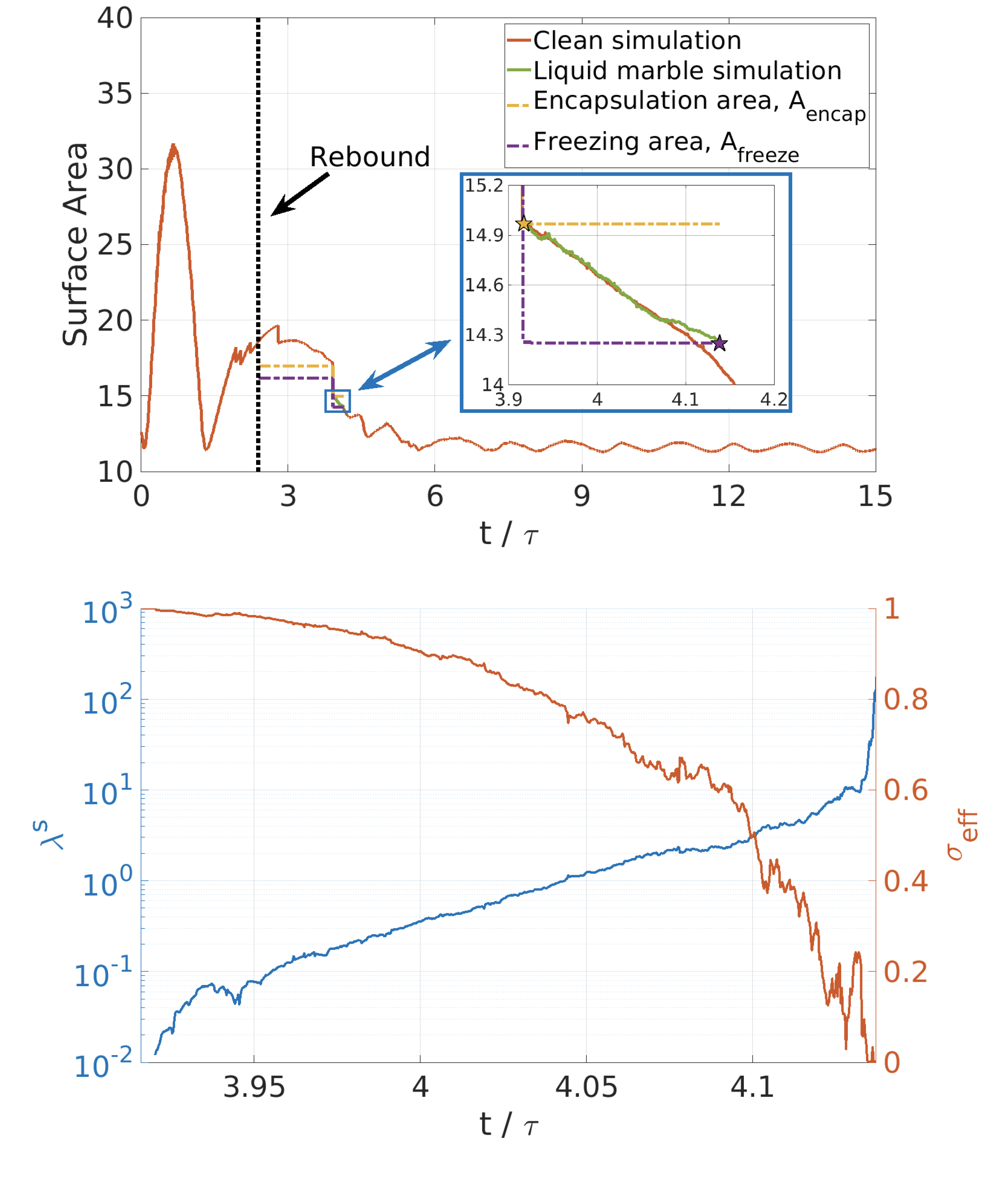} 	
	\end{center}    \caption{Plots for the We = 71 simulation. (Top panel) Surface area for the clean simulation with overlaid surface viscous simulation post-encapsulation, with an inset focusing on the period between the moment of encapsulation and (deformed) liquid marble formation in the surface viscous simulation. The encapsulation and freezing areas reduce following reductions in primary drop surface area due to satellite drop ejections in the surface viscous simulation. (Bottom panel) Dilatational surface viscous parameter $\lambda^s$ and effective surface tension $\sigma_{\text{eff}}$ from the moment of encapsulation to the moment of deformed liquid marble formation.}
    \label{FIG:We=71_SurfArea_and_lambda&sigma}
\end{figure}

\begin{figure}
	\begin{center}
        	\includegraphics[width=\textwidth*19/20]{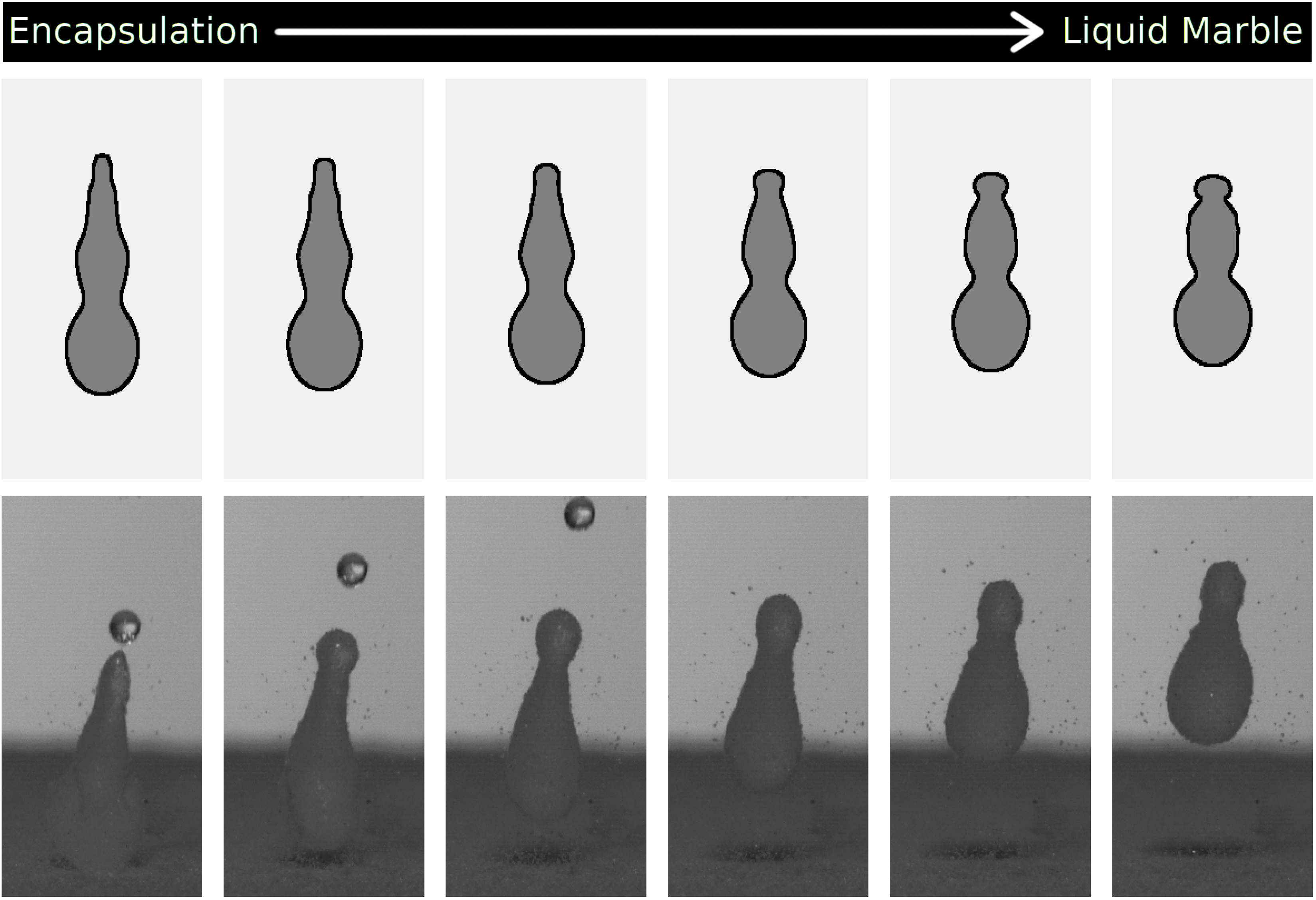} 	
	\end{center}	
	\caption{Drop profile comparison from the moment of encapsulation to the moment of deformed liquid marble formation between the simulation with We = 71 and $\gamma = 2.66$, and a powder bed experiment with We = 78 and $\gamma = 2.69$. Encapsulation is on the far left, with the deformed liquid marble on the far right, and intermediary images between them. }
    \label{FIG:LiquidMarbles_We=71_Dynamics}
\end{figure}

\begin{flushleft}
\emph{We=71}
\end{flushleft}
As shown in Figure \ref{FIG:We=71_SurfArea_and_lambda&sigma}(top), the difference in initial encapsulation area and initial freezing area is very small in this case, and for the short time between encapsulation and liquid marble formation, the difference in surface area between the clean and surface viscous case is \emph{not} significant. In Figure \ref{FIG:We=71_SurfArea_and_lambda&sigma}(bottom) a steady increase in $\lambda^s$ is matched by a steady \emph{decrease} in the effective surface tension towards zero, reaching it shortly before $\lambda^s$ diverges and the freezing area is reached. If we look at the drop shapes for this simulation in Figure \ref{FIG:LiquidMarbles_We=71_Dynamics}, we see little change between the drop at encapsulation and liquid marble formation; the retraction of the drop apex into the rest of the drop is sufficient to close the gap between the encapsulation and freezing areas. Also shown in this figure is an example from a We = 78 ($\gamma = 2.69$) experiment similarly showing that the short retraction of the drop apex into the remainder of the drop is sufficient to freeze an encapsulated drop.

When the encapsulation and freezing areas are as close as in this case (and beyond), the surface viscous model becomes less important in the formation of deformed liquid marbles; what matters here is the threshold we have developed in terms of the parameters $\alpha_{encap}$ and $\alpha_{freeze}$ constructed following observations of experiments (see Figure \ref{FIG:LM_Radius_Alpha}(b)). There is simply not enough time for surface viscosity to have a significant effect on drop dynamics, barring a poor choice for $\beta$ (see Appendix \ref{SEC:Large_Beta}), for the shape of the liquid marble to be drastically altered.

\begin{flushleft}
\emph{We=87}
\end{flushleft}
Here, encapsulation and freezing occur simultaneously, meaning there are \emph{no} surface viscous dynamics, and the shape of the liquid marble is determined entirely by the preceding dynamically-clean drop impact simulation. In terms of our model, this occurs because the concentration of the adsorbed powder on the drop interface is already at the critical freezing threshold when encapsulation occurs.

\begin{figure}
 	\begin{center}
        	\includegraphics[width=\textwidth*19/20]{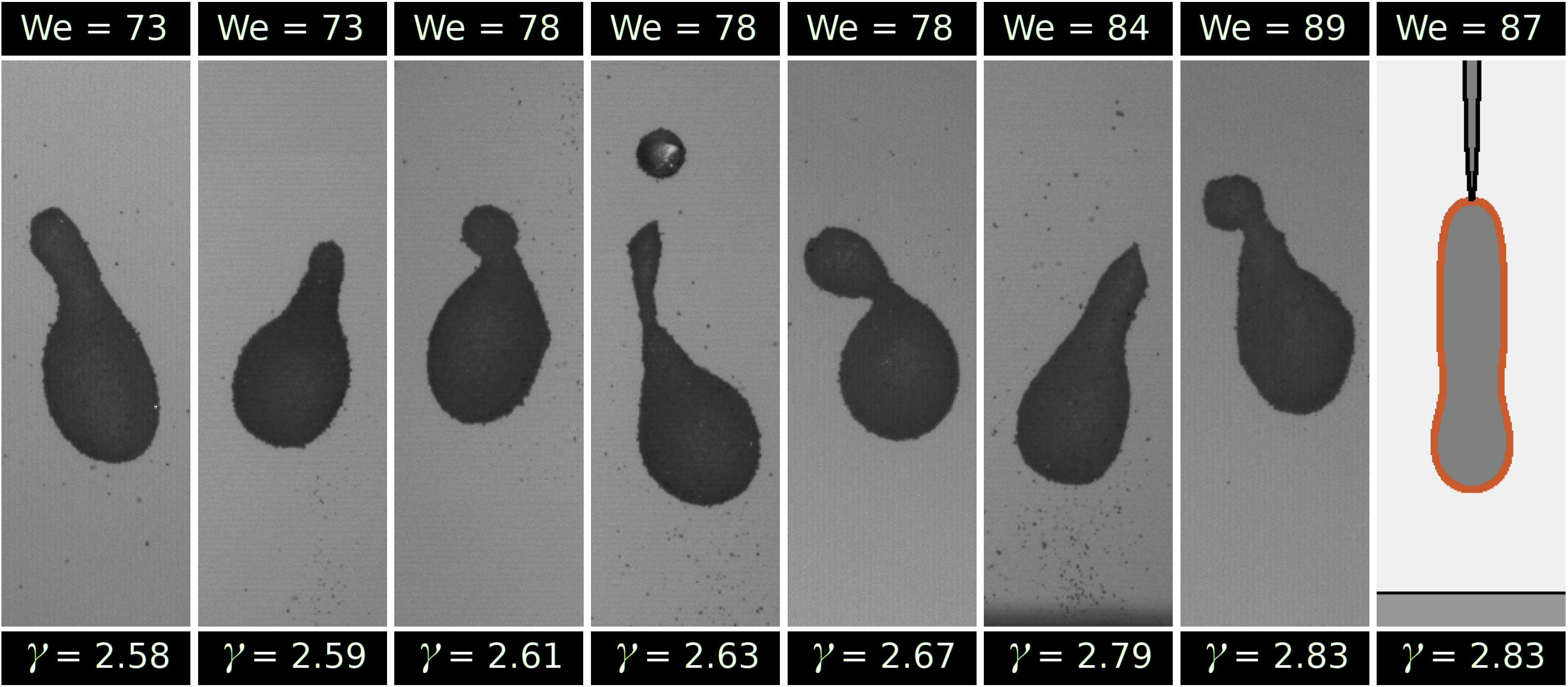} 	
	\end{center}    \caption{Examples of deformed liquid marbles created at the moment of encapsulation in powder bed impact experiments, and the We = 87 ($\gamma = 2.83$) simulation. The deformed liquid marble in the simulation (outlined in red) forms as an immediate result of a satellite drop ejection.}
    \label{FIG:LiquidMarbles_InstantMarbles}
\end{figure}

In Figure \ref{FIG:LiquidMarbles_InstantMarbles} we see our instantaneously-formed liquid marble compared to other such liquid marbles observed in our experiments. There is reasonable agreement between our simulation and these experiments, but there are signs that we are close to the limits of the validity of our model for liquid marble formation, which are discussed here.

\subsubsection{Limitations of the Model at higher We}

For the high-We experiments shown in Figure \ref{FIG:LiquidMarbles_InstantMarbles}, the liquid marbles are seen to be `bottom heavy', with apparent \emph{local jamming} of the interface prior to encapsulation (before the drop apex has been coated in powder), with the motion of the bottom of the drops being similar to that of a rigid body translation. This property of the surface dynamics at high We is \emph{not} present in our simulations, which is because there can be only \emph{global}-jamming (with the entire drop forming a liquid marble) in our model, rather than distinct regions freezing at different times. Interestingly, recall (see Figure \ref{FIG:LM_Radius_Alpha}) that the extrapolated value of $\alpha_{encap}<\alpha_{freeze}$ for this particular simulation (We = 87, $\gamma = 2.83$), which could be an indication that the powder coated region of the drop interface reaches the critical freezing threshold for powder concentration prior to encapsulation., i.e. the local-jamming effect as observed in the experiments.

Another point of discrepancy is our inability to perfectly recreate the pre-encapsulation rebound dynamics for drops impacting onto superhydrophobic powder beds, partly due to our inability to capture the energy loss associated with the interactions between the drop and powder while on the substrate, which is particularly significant at higher We. For the We = 87 ($\gamma = 2.83$) simulation, although the shape at encapsulation/freezing does resemble the drops from experiments in Figure \ref{FIG:LiquidMarbles_InstantMarbles}, it does so due to the fortunate ejection of a particularly long satellite droplet, and forms a more cylindrical shape than any of the experiments. Furthermore, our drop impact model is axisymmetric, and for higher impact Weber numbers, drops exhibit minor breaks in this symmetry which can affect retraction and rebound dynamics, and cannot be captured with our current model. 

In summary, the substrate deformation and local jamming of the drop interface limits our model's capability to describe liquid marble formation, and no comparisons for higher We are made.

\section{Discussion}{\label{SEC:Discussion}}

Drop impact experiments were conducted on superhydrophobic powder beds to better understand the process of liquid marble formation, specifically to aid in the development of the first computational model to numerically simulate this process. Drop impacts were also conducted onto rigid impermeable superhydrophobic substrates to motivate simplifications to the mathematical modelling.

From the impact experiments, we have identified a novel relationship between the maximum spreading diameter of an impacting drop on a superhydrophobic powder bed, and the surface area of the rebounded drop at the moment of encapsulation and deformed liquid marble formation. Using numerical simulations of drop impact on a rigid impermeable superhydrophobic substrate, this relationship was appropriately extended to the maximum contact area between the drop and powder bed to relate the greatest extent of powder coverage to the conditions for the aforementioned critical events.

The first reporting of a computational model for liquid marble formation via drop impact is given, motivated by our experiments, and utilising previously reported surface viscous properties of particle-laden interfaces at high concentrations. The model incorporates dilatational surface viscous effects using the Boussinesq-Scriven constitutive law and is included in the model equations via an effective surface tension, with the important (and unexpected) consequence of surface viscous forces only becoming incorporated following encapsulation of the drop. The strength of the surface viscous effects are chosen by quantitative comparison to oscillatory decay in experiments. In general terms, the events transpiring in simulations were of drop impacts with encapsulation followed by either: (i) a \emph{slow} decay of drop oscillations, (ii) a \emph{rapid} decay of drop oscillations with a spherical liquid marble formed prior to the drop landing back on the substrate, or (iii) the freezing of the drop interface in which a deformed liquid marble is created. The drop shapes at encapsulation and liquid marble formation in simulations, as well as post-encapsulation dynamics, were found to match well qualitatively and semi-quantitatively with experiments.

What has been shown is the simplest model for the liquid marble formation process by drop impact, and is a valuable starting point for future research. Avenues of interest include (i) particle-based simulations to gain a greater insight into the rheology of deformed liquid marbles, (ii) the incorporation of substrate {\color{black} deformation to better match} the drop shapes when rebounding from the substrate, (iii) permitting shear surface viscosity to provide additional damping and which will also lead to localised damping effects due to the right hand side of (an equivalent of) the tangential equation in (\ref{EQN:ExpandedSurfaceStress}) being nonzero, and (iv) lifting the assumption of constant powder concentration in order to describe local jamming effects.

\section{Acknowledgements}
We acknowledge financial support by EPSRC (Grants No. EP/N016602/1, No. EP/S029966/1, No. EP/P031684/1, and No. EP/HO23364/1)

\bibliographystyle{unsrt}
\bibliography{liquid_marble_paper}

\appendix

\section{Image Analysis}{\label{APP:Image_Analysis}}
Following completion of experiments, image analysis is conducted to provide a variety of measurements; most importantly for the maximum spreading diameter and an approximation of drop surface area. The maximum spreading diameter is easy to measure as it only requires pin-pointing the extreme left- and right-most pixels of the drop at maximum extension, whereas other quantities rely on data that is harder to obtain, such as the shape of the drop boundary.

\begin{figure}[H]
	\begin{center}
        	\includegraphics[width=\textwidth*19/20]{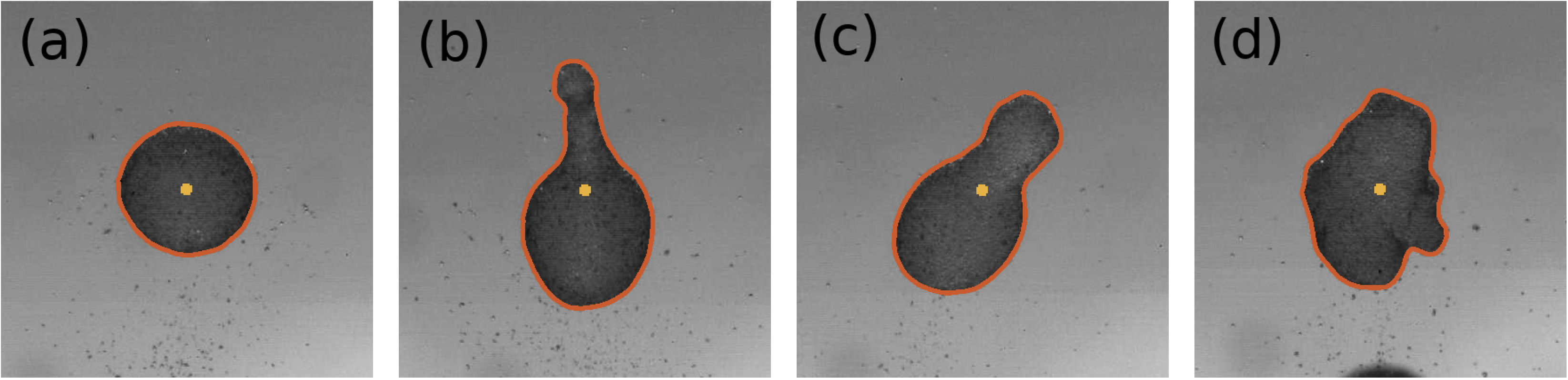} 	
	\end{center}
    \caption{Liquid marbles with overlaid smooth boundaries and centroids. The impact Weber number for each experiment is: (a) We = 49, (b) We = 59, (c) We = 69, (d) We = 79.}
    {\label{FIG:ImageAnalysis_Outlines}}
\end{figure}

To obtain an outline of the shape for a coated drop, a frame is taken from an experimental video and the contrast is altered so that the drop profile appears as a black mass of pixels on a white background. The pixels at the boundary of the drop are then taken and ordered sequentially. As this ordered boundary data is taken directly from pixels, the boundary is not smooth; which can introduce spurious results when calculating quantities such as surface area (in fact this gets worse as multiple layers of powder adhere to the drop interface). We therefore smooth the boundary data, and do so by applying Savitzky-Golay filtering \cite{savitzky}; in our case by fitting successive subsets of 25 adjacent data points that constitute the unfiltered drop boundary (consisting of 400-500 data points in total) with third-order polynomials. Figure \ref{FIG:ImageAnalysis_Outlines} shows images of liquid marbles overlaid with smoothed boundaries, constructed using the above process, along with identification of their centroids (that is, the two dimensional centre of mass).

To obtain an approximation of drop surface area with just one camera view, we have to assume that the images we see from experiments represent an axisymmetric shape, which appears to be the case (in varying degrees) for all but the highest impact Weber numbers, where the drop experiences splashing or the fingering instability (where the rim of the spreading drop splits into liquid `fingers'). Approximating the surface area under an axisymmetry assumption requires the choosing of an appropriate axis of symmetry. Prior to impact, the axis of symmetry is the vertical line that traces the drop's descent onto the substrate, so assuming axisymmetry is maintained during the impact and after rebound, this axis will be unchanged throughout the experiment. We do observe however that liquid marbles will often rotate in the air after forming, likely caused by heterogeneity of the powder bed, but fortunately the drops still appear to maintain reasonable axisymmetry about a now-rotated axis.

\begin{figure}
	\begin{center}
        	\includegraphics[width=\textwidth*19/20]{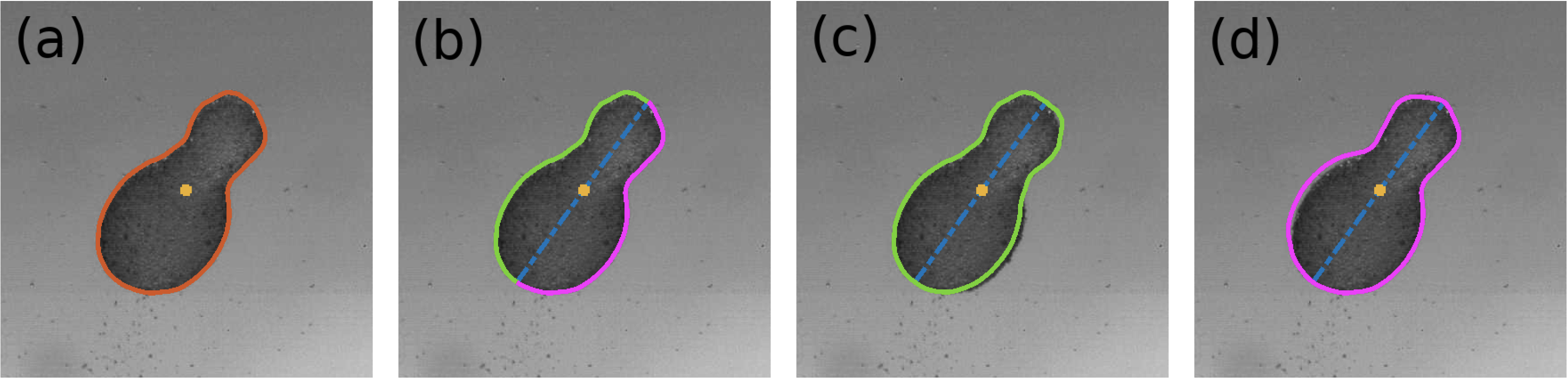} 	
	\end{center}
    \caption{Example of the process of approximating the surface area of a liquid marble: (a) Identify the boundary (red) of the two dimensional image of the drop and calculate the centroid of the filled shape, (b) Draw a line (blue) through the centroid that can reasonably act as an axis of symmetry for the drop shape - this splits the boundary into a left and right component (green and pink), (c-d) Construct a surface-of-revolution using the left and right components of the drop boundary. The `true' surface area is approximated as the average of the approximations in (c) and (d).}
    {\label{FIG:ImageAnalysis_AxisOfSymmetry}}
\end{figure}

For our approximation, we choose an axis of symmetry by drawing a straight line connecting two points of the drop boundary such that this line passes through the drop centroid, and that the orientation of this line matches closely to the visual evolution of the axis in an experiment (see example in Figure \ref{FIG:ImageAnalysis_AxisOfSymmetry}). This axis splits the drop shape into two components, referred to as left and right components. Two approximations are then made for the drop surface area by assuming each component, when made into a surface-of-revolution, is representative of the three dimensional drop shape. The surface area for the liquid marble, $A_{LM}$, is then taken as the average of these two approximations, that is,
\begin{equation}
	A_{LM} = \frac{1}{2} \left(2\pi\int_{\text{left}} r \ \text{d}s \ + \ 2\pi\int_{\text{right}} r \ \text{d}s\right),
\end{equation}
where d$s$ denotes the line element along the drop boundary in the left or right component. Ultimately, we find that after applying this averaging, surface area calculations are not particularly sensitive to the precise placement of the axis of symmetry, as long as a `sensible' choice is made.

\section{Derivation of the Initial Freezing Area}{\label{APP:Initial_Freezing_Area}}

Suppose that a drop experiences $n$ satellite drop ejections that remove powder from the primary drop before eventually the drop freezes due to the powder concentration reaching the freezing threshold $c_{freeze}$. Let $M_i^{(-)}$ and $M_i^{(+)}$ denote the mass of adsorbed powder on the primary drop interface immediately before and after the $i$-th pinch-off event, respectively, for $i = 1,...,n$. Let $A_i^{(-)}$ and $A_i^{(+)}$ denote the powder coated surface area of the primary drop for the same event, respectively. Finally, let $M_{final}$ and $A_{final}$ denote the adsorbed powder mass, and surface area of the primary drop at the moment of freezing some time after all of these satellite drop ejections, respectively.

As powder mass is only lost at discrete pinch-off events, we can write the powder mass at the moment of eventual freezing as the original mass multiplied by the proportion of mass still remaining after each ejection, that is,
\begin{equation}
    M_{final} = M_{contact} \cdot \frac{M_1^{(+)}}{M_1^{(-)}} \cdot \frac{M_2^{(+)}}{M_2^{(-)}} \cdot \ldots \cdot \frac{M_n^{(+)}}{M_n^{(-)}}.
\end{equation}
Using that $M = cA$, we can rewrite this equation as
\begin{equation}
    c_{freeze} A_{final} = M_{contact} \cdot \frac{A_1^{(+)}}{A_1^{(-)}} \cdot \frac{A_2^{(+)}}{A_2^{(-)}} \cdot \ldots \cdot \frac{A_n^{(+)}}{A_n^{(-)}},
\end{equation}
where we have used the fact that powder concentration is continuous through ejection events. Dividing both sides through by $c_{freeze}$ and using $A_{freeze}^{(0)} = M_{contact}/c_{freeze}$ then gives us
\begin{equation*}
    A_{final} = A_{freeze}^{(0)} \cdot \frac{A_1^{(+)}}{A_1^{(-)}} \cdot \frac{A_2^{(+)}}{A_2^{(-)}} \cdot \ldots \cdot \frac{A_n^{(+)}}{A_n^{(-)}},
\end{equation*}
or equivalently,
\begin{equation}{\label{EQN:A_freeze_computable}}
    A_{freeze}^{(0)} = A_{final} \cdot \frac{A_1^{(-)}}{A_1^{(+)}} \cdot \frac{A_2^{(-)}}{A_2^{(+)}} \cdot \ldots \cdot \frac{A_n^{(-)}}{A_n^{(+)}},
\end{equation}
where the right hand side of (\ref{EQN:A_freeze_computable}) is made up exclusively of quantities that are directly measurable from experiments. Importantly, the surface area immediately following a pinch-off event is \emph{not} the same as the surface area before the next pinch-off event (for example, $A_1^{(+)} \neq A_2^{(-)}$) because of dynamics causing surface area change between these events. Therefore (\ref{EQN:A_freeze_computable}) allows us to calculate, for any experiment that produces a liquid marble, a value of the initial freezing area $A_{freeze}^{(0)}$, that is, the critical area threshold for interfacial freezing that would be observed if no adsorbed powder mass was removed from the primary drop.

\section{Time-dependent Encapsulation Area and Freezing Area}{\label{APP:TimeDep_Encap_Freeze}}

Following from Appendix \ref{APP:Initial_Freezing_Area}, powder concentration is continuous through pinch-off events. To determine the values of the primary drop surface area associated with the (spatially constant) powder concentration being equal to its value at encapsulation ($c_{LG}$) and at freezing ($c_{freeze}$) over time, the time dependent encapsulation area $A_{encap}(t)$ and freezing area $A_{freeze}(t)$ must change to account for powder and surface area loss through pinch-off events.

Consequently, the following equation notes how the encapsulation area and freezing area change through the $j$-th pinch-off event associated with powder loss on the primary drop:
\begin{equation}{\label{EQN:Area_Enc_time_dependent}}
    A_{encap}(t_j^+) = A_{encap}(t_j^-) \cdot \frac{A_p(t_j^+)}{A_p(t_j^-)} \quad \text{and} \quad A_{freeze}(t_j^+) = A_{freeze}(t_j^-) \cdot \frac{A_p(t_j^+)}{A_p(t_j^-)},
\end{equation}
where $t_j^-$ and $t_j^+$ denote the times immediately before and after the pinch-off event, and $A_p(t)$ denotes the surface area of the powder coated region of the primary drop at time $t$. For example, if encapsulation occurs due to a pinch-off event, then $j = 1$ as all previous pinch-off events were for clean satellite drops, $A_p(t_1^-)$ is the total surface area of all \emph{powder coated} regions prior to pinch-off, and $A_p(t_1^+)$ is the surface area of the primary drop following the pinch-off. For all subsequent pinch-off events, $A_p(t_j^-)$ and $A_p(t_j^+)$ denote the total surface area of \emph{all} regions of the primary drop prior to, and following, a pinch-off event respectively, as there are no regions without a powder coating.

The encapsulation area $A_{encap}(t)$ and freezing area $A_{freeze}(t)$ are therefore piecewise constant in time, reducing only at discrete pinch-off events where powder mass is lost from the primary drop surface. Up to the first pinch-off event that reduces powder mass on the primary drop surface, we have $A_{encap}(t) = A_{encap}^{(0)}$ and $A_{freeze}(t) = A_{freeze}^{(0)}$.

\section{Effect of large $\bar{\beta}$}
\begin{figure}[h]
	\begin{center}
        	\includegraphics[width=\textwidth*19/20]{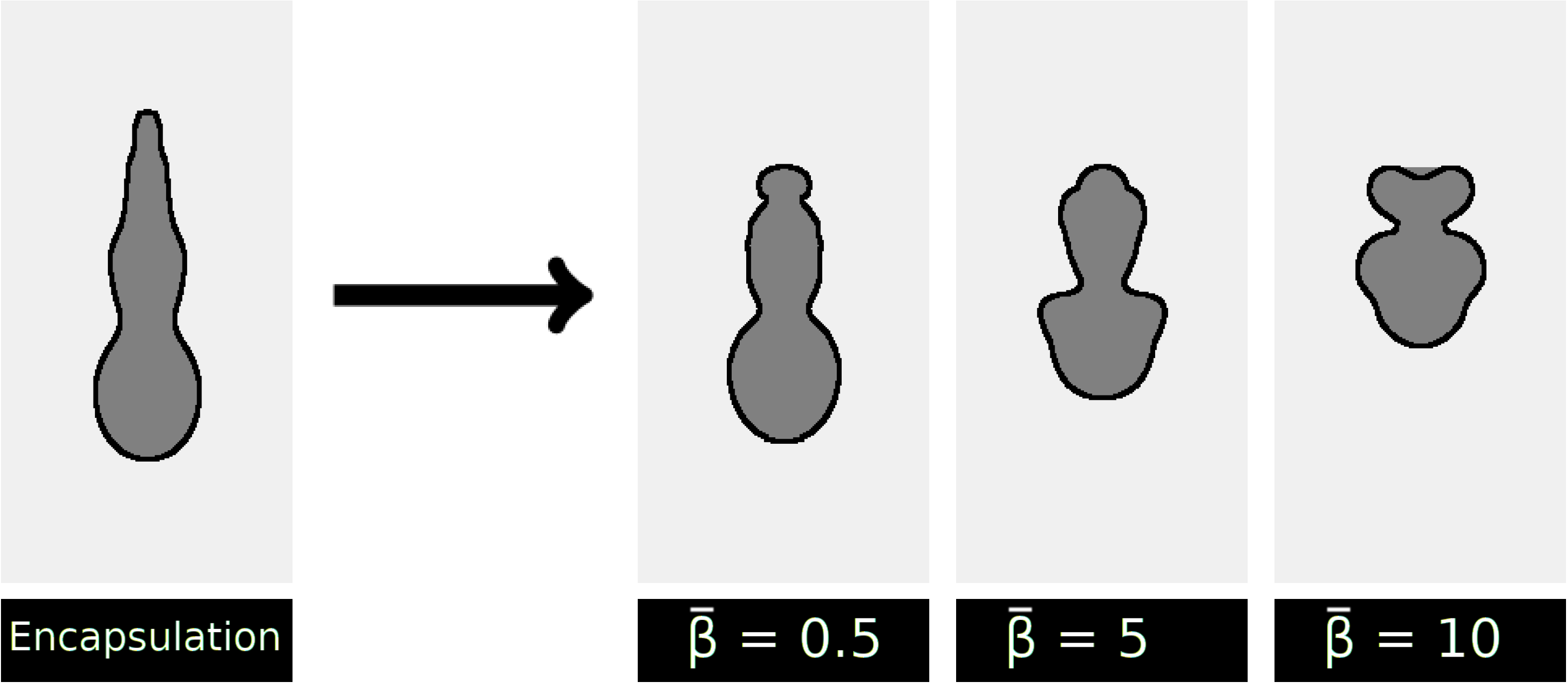} 	
	\end{center}
    \caption{Drop profiles showing the drop at encapsulation for We = 71 ($\gamma = 2.66$) on the left; the remaining images show the liquid marble created using $\bar{\beta} = 0.5$, $\bar{\beta} = 5$, and $\bar{\beta} = 10$.}
    \label{FIG:LiquidMarbles_StrangeMarbles}
\end{figure}

{\label{SEC:Large_Beta}}
As an extra assurance that our choice of $\bar{\beta}= 0.5$ in \S\ref{SEC:ChoosingBeta} is reasonable, we see in Figure \ref{FIG:LiquidMarbles_StrangeMarbles} the drop at encapsulation compared to a set of liquid marbles formed with $\bar{\beta} = 0.5, 5, 10$ for We = 71 ($\gamma = 2.66$). As discussed in the previous section and shown in Figure \ref{FIG:LiquidMarbles_DeformedMarbles}, the liquid marble formed with $\bar{\beta} = 0.5$ is similar to those observed in experiments for similar impact Weber numbers and spreading factors. However, for the $\bar{\beta} = 5$ and $\bar{\beta} = 10$ cases, the liquid marbles are far removed from those seen in experiments for this impact Weber number (and those similar to it), where the formed liquid marbles typically differ \emph{little} from the shapes at encapsulation.

\end{document}